%% file: bare_conf_compsoc.tex
\documentclass[conference,compsoc]{IEEEtran}
\IEEEoverridecommandlockouts
\ifCLASSOPTIONcompsoc
  \usepackage[nocompress]{cite}
\else
  \usepackage{cite}
\fi
\ifCLASSINFOpdf
\else
\fi
\ifCLASSOPTIONcompsoc
 \usepackage[caption=false,font=footnotesize,labelfont=sf,textfont=sf]{subfig}
\else
 \usepackage[caption=false,font=footnotesize]{subfig}
\fi

\usepackage{multirow}
\usepackage{makecell}
\usepackage{pgfplots}
\usepackage{float}
\usepackage{hyperref}
\hypersetup{colorlinks=true, urlcolor=blue, citecolor=black, linkcolor=black}

\begin{document}
%
\title{High-resolution Image-based Malware Classification \\using Multiple Instance Learning}

\author{\IEEEauthorblockN{Tim Peters}
\IEEEauthorblockA{University of Southampton\\United Kingdom\\
tpp1u22@soton.ac.uk}
\and
\IEEEauthorblockN{Hikmat Farhat}
\IEEEauthorblockA{University of Southampton\\United Kingdom\\
h.farhat@soton.ac.uk}
}


%


\maketitle

\begin{abstract}
This paper proposes a novel method of classifying malware into families using high-resolution greyscale images and multiple instance learning to overcome adversarial binary enlargement. Current methods of visualisation-based malware classification largely rely on lossy transformations of inputs such as resizing to handle the large, variable-sized images. Through empirical analysis and experimentation, it is shown that these approaches cause crucial information loss that can be exploited. The proposed solution divides the images into patches and uses embedding-based multiple instance learning with a convolutional neural network and an attention aggregation function for classification. The implementation is evaluated on the Microsoft Malware Classification dataset and achieves accuracies of up to 96.6\% on adversarially enlarged samples compared to the baseline of 22.8\%. The Python code is available online at \url{https://github.com/timppeters/MIL-Malware-Images}.
\end{abstract}


%
\IEEEpeerreviewmaketitle

\section{Introduction}
\input{sections/1_introduction}

\section{Background}
\input{sections/2_background}

\section{Methodology}
\input{sections/3_methodology}

\section{Results and Discussion}
\input{sections/4_results}

\section{Limitations}
\input{sections/5_limitations}

\section{Conclusion}
\input{sections/6_conclusion}






%


\bibliographystyle{IEEEtran}
\bibliography{references}

\end{document}

%% file: sections/1_introduction.tex
Being able to classify malware samples as benign or malicious is vital to protect devices from being infected. Due to the frequent re-use of source code, malware can be further categorised into families. Developing techniques for classifying samples into known families is an important area of research, as family labels can provide insights into the behaviour and origin of malware samples and aid in remediation efforts \cite{Joyce2021ALabels}.

Machine learning methods to classify malware using images have been widely proven to be successful \cite{Nataraj2011MalwareImages, Bijitha2022OnTechniques}. They involve converting malware samples into images (byteplots) and subsequently use machine learning to perform image classification. There are numerous benefits to image-based malware classification, including clear visual similarities among malware samples from the same family \cite{Nataraj2011MalwareImages}, no requirement for disassembly or execution of samples, obfuscation resistance  \cite{Marastoni2018ASimilarity}, and platform independence.

A crucial downside, however, is that the images can be very large, with a 4MiB malware sample mapping to a 2048×2048 pixel byteplot. Therefore, previous research using byteplots and deep neural networks have scaled the images down in order to achieve reasonable speed and memory usage. This resizing leads to information loss that adversaries could exploit. In this paper, we show that the drastic re-sizing of images allows for malware to reduce the accuracy of these models by adding vast amounts of redundant data to the original sample, which ``dilutes" the byte information related to the malware when the pixels are interpolated. In order to overcome this type of adversarial attack, a model would need to consider the whole binary sample. Otherwise, the exact byte values could be manipulated.

Our proposed solution uses Multiple Instance Learning (MIL), a paradigm where each large image is divided into a bag of smaller image patches, and one label is associated with the whole bag \cite{Mandlik2021MappingNetworks}. The bag-level label is learned with the instance-level data. This approach retains all the data in the image, while aiming to maintain the benefits of image-based analysis.

%% file: sections/2_background.tex
\subsection{Image-Based Malware Analysis}
\label{subsec:Image-Based Malware Analysis}

Nataraj et al. \cite{Nataraj2011MalwareImages} introduced the idea of a ``malware image", where the binary data of a malware sample is converted into an 2D image called a byteplot. This is accomplished by reading the binary data as a 1D vector of 8-bit unsigned integers, which can then be reshaped into a 2D matrix and visualised as a greyscale image with pixel values between 0-255. This converts malware classification into an image recognition problem, as the resulting bitmap images can be classified using conventional computer vision techniques.

\begin{figure*}[!t]
	\centering
	\includegraphics[width=0.8\textwidth]{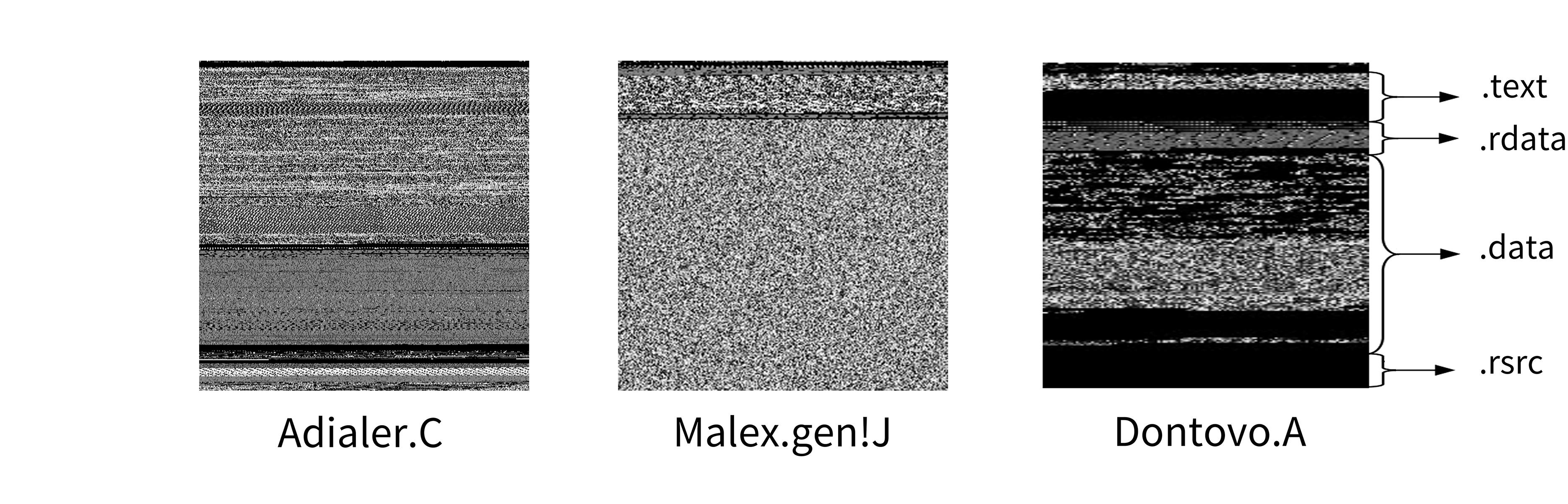}
	\caption{Examples of malware byteplots for three different malware families, clearly showing structural differences. Adapted from \cite{Nataraj2011MalwareImages}.}
	\label{img:malware_images_families}
\end{figure*}

By visualising malware as an image, the authors observed that there were significant similarities in texture and layout of malware belonging to the same family, as shown in Figure \ref{img:malware_images_families}. In contrast, images of malicious and benign samples or samples belonging to different families appeared to be distinct. This is likely due to the fact that most malware variants are created via small modifications of a previous sample to change the signature enough to avoid detection. As computer vision techniques for classification are generally designed to be insensitive to small perturbations in the input, they are appropriate for this task \cite{Saridou2022SAGMADASets}. Even with obfuscation techniques such as packing and encryption that modify the entire sample, malware families tend to use custom obfuscators, leading to structural similarities between variants \cite{Conti2022AVisualization, Nataraj2011MalwareImages, Nataraj2011AAnalysis}. Many image-based malware classification methods have shown good performance for both malware family classification \cite{Vasan2020IMCFN:Architecture, Nataraj2011MalwareImages, Bijitha2022OnTechniques} and malware detection (malicious/benign) \cite{Saridou2022SAGMADASets, Chen2020STAMINA:Classification}.

Unlike other static and dynamic detection techniques, classifying malware in this way relies only on the raw bytes. This means that no disassembly or execution of the sample is required, making it very fast and scalable \cite{Chen2020STAMINA:Classification}.

\subsection{Problems with Resizing}
\label{subsec:Problems with Resizing}

Classifying images using a standard fully-connected neural network is not feasible, as $n$ input pixels and $n$ units in the first hidden layer would already require $n^2$ weights. Convolutional Neural Networks (CNNs) are often used instead, as due to their local receptive fields and weight sharing, the number of weights in the convolutional layers is independent of the image size. The output of a convolutional layer, however, \textit{is} dependent on the input size, the number of kernels, and the kernel stride \cite{Russell2020ArtificialEdition}. Although requiring a lot less memory, the intermediate representations and fully-connected layers still depend on the input size, and as such it is not feasible to directly use very large images with CNNs.

Additionally, due to the fully-connected layers, the output of the last convolutional layer must be a fixed size. The most common approach to achieve this is by having fixed-size input images that are created by resizing or cropping images during the pre-processing stage. The majority of models therefore use input image sizes of 224×224px, 256×256px, or smaller \cite{Szegedy2015RethinkingVision, Tan2019EfficientNet:Networks, He2015DeepRecognition}. In most applications, the images used are natural and can be downsampled without losing crucial information. Therefore, CNNs still show remarkable performance for many image classification tasks \cite{Rawat2017DeepReview, Sharma2018AnClassification}.

A major problem with classifying malware using raw bytes is the size of the sample. Not only do the size of samples vary widely, which can be problematic for many machine learning algorithms, but images generated from malware samples are typically quite large. For example, a 10MB sample would result in a  $\sim$3162×3162px greyscale image. To deal with this issue, most studies from the literature downsample the generated images to a fixed size, such as 128×128px or 224×224px. As these images are not natural, resizing them is not numerically meaningful as exact byte values are interpolated \cite{Raff2017MalwareEXE}. Despite this, researchers rarely acknowledge the crucial data loss that this causes, only noting that ``some important features" are lost \cite{Vasan2020IMCFN:Architecture}, that the network is deep enough to still extract features \cite{Chen2020STAMINA:Classification}, or that the information needed for classification is not reduced \cite{Marastoni2021DataContext}. The reason for their neglect is likely that they still produce accurate and robust results after downsampling their inputs. Given that working with smaller images makes the training process significantly faster, it is still a useful classification method for triaging conventional malware.

The problem, however, is that the loss of information can be exploited by an adversary by adding vast amounts of redundant data to the sample, increasing the image resolution drastically. If the ratio of redundant to useful information in the sample is extreme enough, the information crucial to classifying the malware would be interpolated away.

Purposefully inflating the size of malware to avoid analysis or detection is a known MITRE ATT\&CK technique \cite{MITRE2020ObfuscatedATTCK}. Not only does it reduce the chance of being collected for further analysis, but many security tools have file size limitations and are therefore not able to handle this type of malware. Mitigation is hard and could require significant changes to the tools as intrinsic limitations of the algorithms are being exploited. Although not extremely common, this technique has been used by threat actors many times, likely due to the common practice of setting file size limits in YARA rules \cite{MITRE2020ObfuscatedATTCK, Ishimaru2017OldData} and YARA's own file size limit of 64MB \cite{YARASupportFileSize}.

\begin{table*}[!t]
\renewcommand{\arraystretch}{1.3}
\caption{The metrics comparing the original image and the resized image for sample ID 0A32eTdBKayjCWhZqDOQ.}
\label{table:lollipop_loss_metrics}
\centering
\begin{tabular}{cccccccccc}
\hline
 Family & Size & Image Size & MSE & SSIM & SNR & Image Entropy & Resized Image Entropy & MI & MI \% of original \\
\hline
    Lollipop & 1163KB & 1091x1091 & 88.3 & 0.207 & -1.29 dB & 6.26 & 5.69 & 0.913 & 14.6\%  \\
\hline
\end{tabular}
\end{table*}

\begin{figure*}[!t]
    \centerline{
    \subfloat[\centering Original]{{\includegraphics[width=0.25\textwidth]{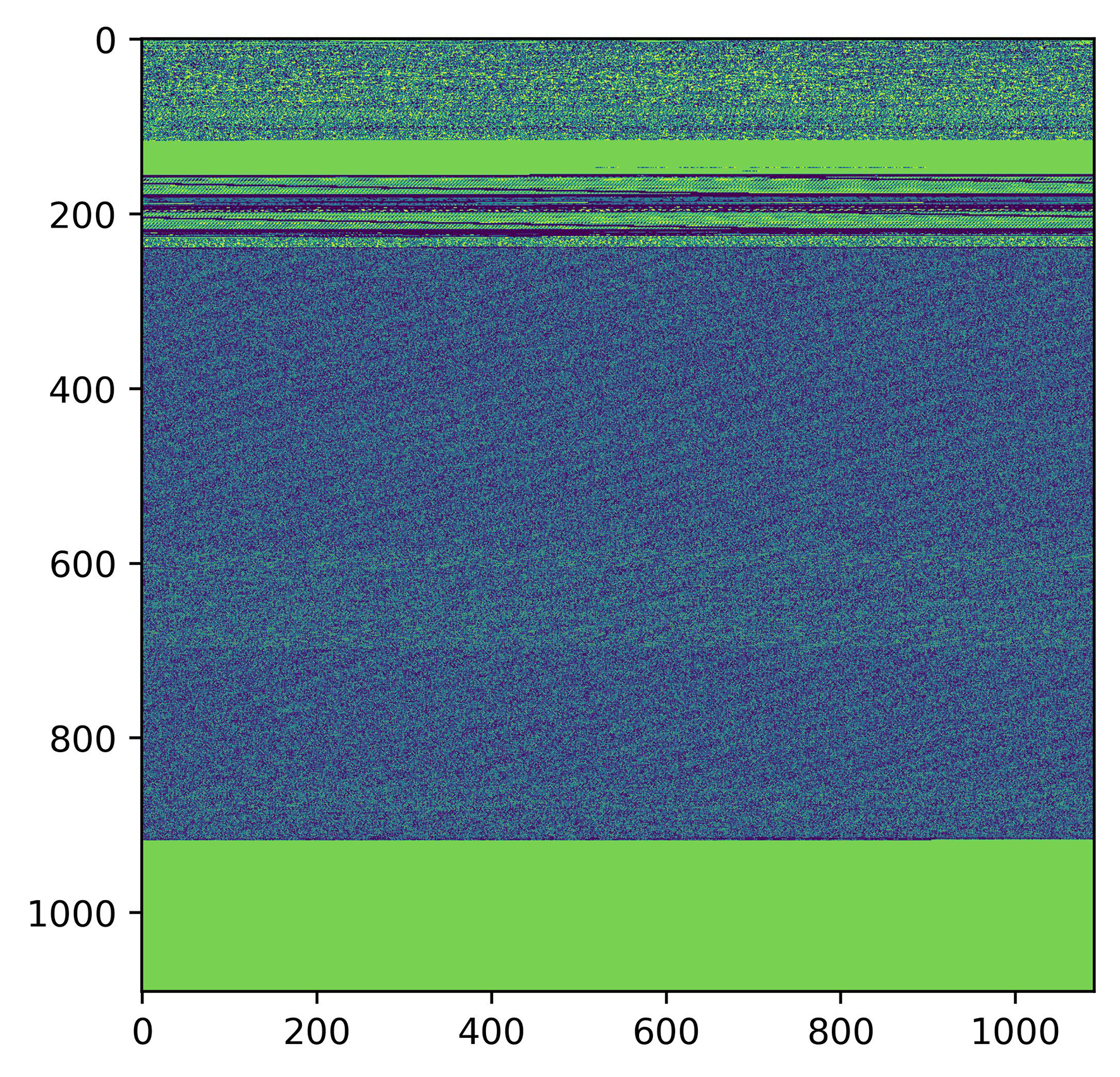} }}%
    \subfloat[\centering Resized]{{\includegraphics[width=0.25\textwidth]{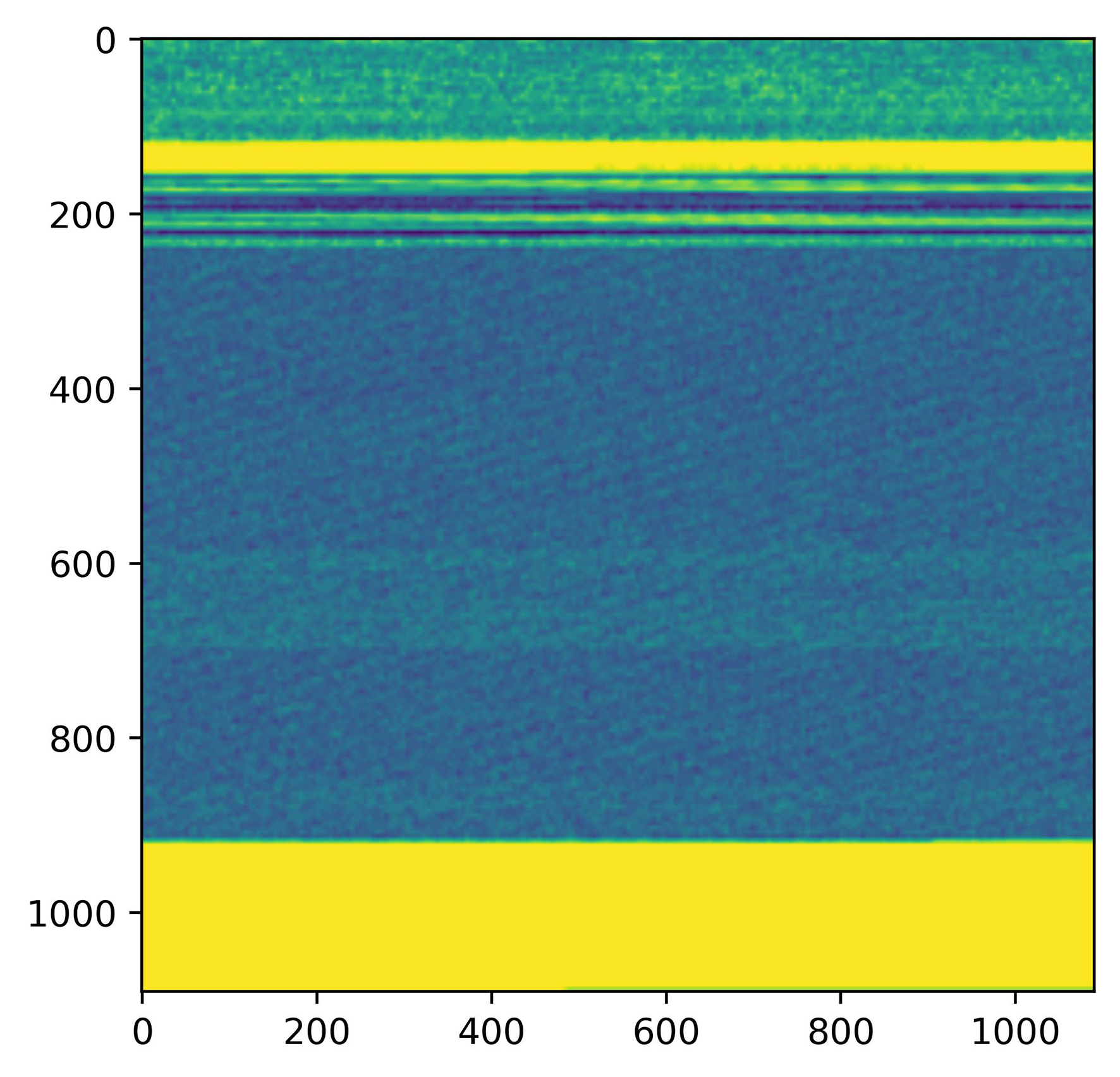} }}%
    \subfloat[\centering Difference]{{\includegraphics[width=0.28\textwidth]{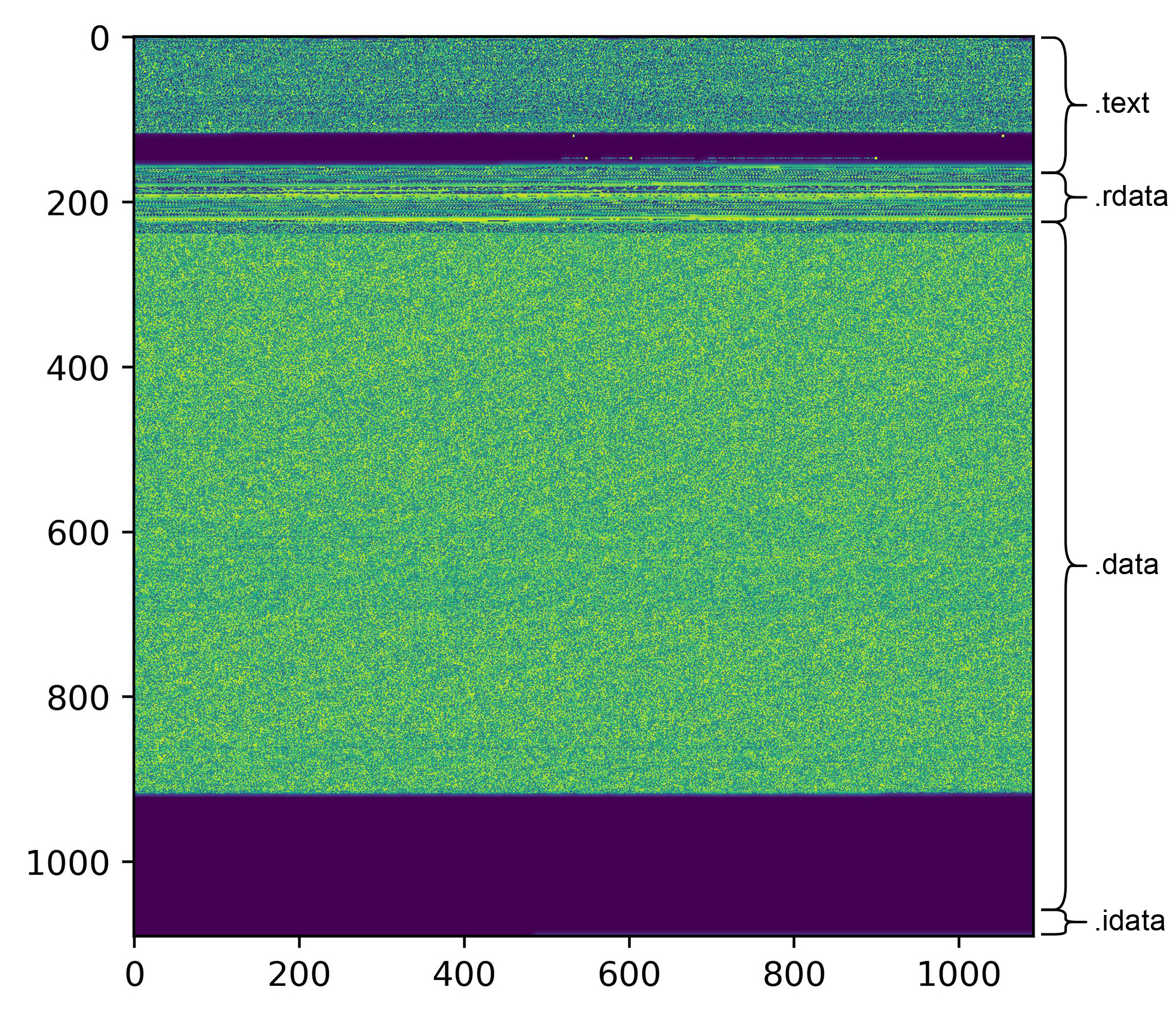} \label{fig:information_loss_diff}}}%
    }
    \caption{From left to right: the original malware image, the downsampled image resized back to the original size, and the difference between the two. A colour map is applied to improve readability.}%
    \label{fig:information_loss_all_3}%
\end{figure*}


\subsubsection{Formalising Data Loss}

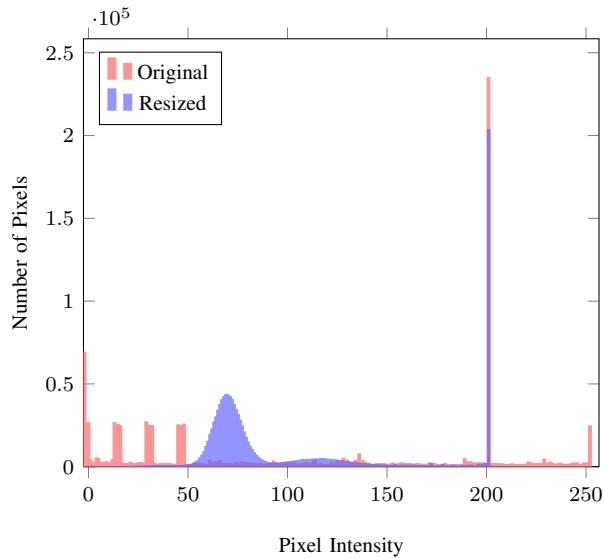
\begin{figure}[!t]
	\centering
	\input{assets/resizing_data_loss/sample_histogram}
	\caption{Pixel intensities for both the original and resized malware image.}
	\label{figure:resizing_loss_histogram}
\end{figure}

To formalise this data loss, we can compare examples of full-resolution malware images to versions that have been downsampled and then resized back to their original size using the commonly used bilinear interpolation algorithm. As an example, a sample is taken from the Microsoft Malware Classification dataset \cite{Ronen2018MicrosoftChallenge} and the following metrics are computed after resizing the image to 224×224 pixels: Mean Squared Error (MSE), Structural Similarity Index (SSIM), Signal-to-Noise Ratio (SNR), and Mutual Information (MI). The difference between the two images and pixel intensity histograms are shown in Figure \ref{fig:information_loss_diff} and Figure \ref{figure:resizing_loss_histogram}, respectively.

The results in Table \ref{table:lollipop_loss_metrics} clearly show that significant information was lost due to the resizing process. A high MSE of 88.3 and a low SSIM of 0.207 both indicate a significant change in image texture. The negative SNR indicates that the power of the noise is greater than the power of the original signal, suggesting that resizing at this scale is causing considerable alterations. Furthermore, the mutual information between the original image and the resized image is only 14.6\% of the information present in the original image, implying data loss.

Figure \ref{fig:information_loss_diff} shows high differences throughout most of the image, but especially parts where there is fine-grained detail, such as the .text and .data sections. This is problematic, considering CNNs trained on raw bytes for Portable Executable (PE) malware classification pay the most attention to these sections \cite{Coull2019ActivationClassification}, as they likely contain the most discriminative information.

The histogram in Figure \ref{figure:resizing_loss_histogram} shows that the distributions of pixel values differ greatly between the two images. Notably, it seems that the bilinear interpolation process causes pixel intensities to cluster around 50-90 in the shape of a normal distribution.

The significant differences between the original malware image and the resized version show that resizing images before classification leads to substantial information loss, much of which would be discriminative of malware or of a malware family, which could crucially reduce classification performance. To illustrate how an attacker could exploit this, the image is artificially enlarged by appending black pixels (null bytes) before resizing and the mutual information of the malware data is calculated (ignoring the padding). 

\begin{figure}[!t]
	\centering
	\input{assets/resizing_data_loss/adversarial_loss}
	\caption{A graph of mutual information (MI) against size of the adversarially enlarged image (original image 1091x1091).}
	\label{figure:resizing_loss_adversarial}
\end{figure}
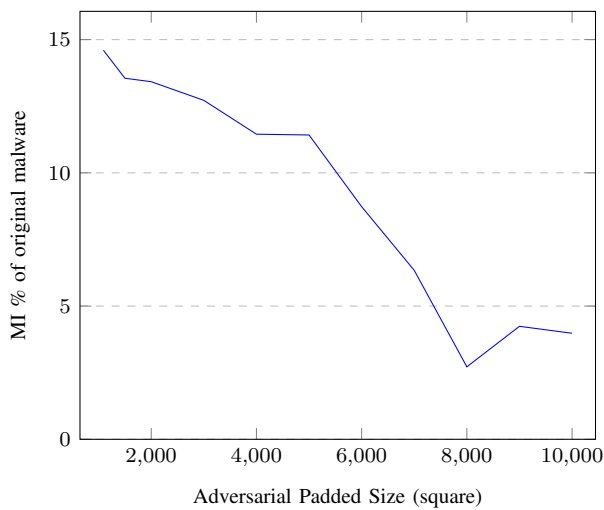

Figure \ref{figure:resizing_loss_adversarial} plots the MI at different amounts of padding, illustrating the information loss relevant to the malware and the effectiveness of this adversarial attack. Clearly, image resizing is not an appropriate pre-processing step for a robust imaged-based malware classifier.

\subsection{Multiple Instance Learning}

To make a classifier robust against adversarial enlargement attacks, image features must be extracted from the original, full-resolution byteplot. There are many traditional computer vision techniques that are able to generate a fixed-size descriptor vector from any image size, such as GIST \cite{Oliva2001ModelingEnvelope} and locality-sensitive image hashing. However, they would still be vulnerable to an adversarial enlargement attack, as the algorithms are not trainable.

Multiple Instance Learning (MIL) can also generate a fixed-size description vector by dividing the input image into fixed-sized patches and calculating a learned weighted average from patch embeddings \cite{Pinckaers2022StreamingImages}. MIL is useful when individual patches (or a combination of them) contain enough information to classify the whole image, such as in histopathology, where small patches of a whole tissue slide can indicate cancer \cite{Couture2018MultipleHistopathology}. Relating this to malware classification, individual patches of a byteplot could contain enough discriminatory information to determine the family, especially when the sample is adversarially enlarged. 

Formally, multiple instance learning is a weakly-supervised machine learning paradigm where each sample $\mathbf{X}$ is represented by a set of N-dimensional vectors $\{\mathbf{x}_1, \mathbf{x}_2, ..., \mathbf{x}_k\}$, where $k$ is the number of instances in the set \cite{Mandlik2021MappingNetworks, Couture2018MultipleHistopathology, Wei2016AnLearning}. The sample is called a bag and the vectors are called instances. The goal is to learn patterns at the instance-level (for which labels may or may not exist) which can be aggregated to predict bag-level labels. We assume that the number of instances per bag $k = |\mathbf{X}_i|$ can vary for different bags. Therefore, a suitable aggregation function should be invariant to the number of instances as well as be permutation-invariant, since no assumptions are made on the ordering or inter-dependence of instances in a bag \cite{Ilse2018Attention-basedLearning, Couture2018MultipleHistopathology}. 

For classifying malware images with MIL, the image (bag) can be divided into patches (instances). The bag level label (malicious/benign or family) is given, but the instance labels are not, as arbitrary parts of an executable cannot be labelled in such a manner. 

One strength of MIL is that it is able to deal with instances having different internal distributions. In fact, it is likely in real-word data that instances in a bag are inherently related to each other and/or drawn from different distributions (non-i.i.d) \cite{Mandlik2021MappingNetworks}. For classifying images with MIL, non-i.i.d is implicitly assumed as the instances are dependent because they come from the same larger image. Furthermore, executable formats such as PE have internal structures with sections having different distributions (e.g. .text and .data), causing patches to be non-i.i.d. Lastly, when dealing with non-i.i.d data, certain instances might be more informative to the bag label than others (e.g. when a sample is adversarially enlarged). For these reasons, MIL could work better for classifying malware byteplots than a tiling-based CNN approach.

Using MIL with CNNs to classify malware bytplots would result in a model that can classify inputs of varying, arbitrary size in a memory efficient manner. However, for the model to be robust against adversarial enlargement attacks, the MIL aggregation function must be robust.

\subsection{Related Work}
In their foundational work, Nataraj et al. \cite{Nataraj2011MalwareImages} proposed the concept of malware visualisation for family classification, based on their observation of visual similarities in texture and layout between samples from the same family. They first converted the malware samples to images with a fixed width based on file size and variable height, and then computed GIST texture features. For classification, they used K-Nearest Neighbours with Euclidean distance. From 10-fold cross validation on a dataset of 9,458 samples from 25 families (Malimg dataset), they obtained an accuracy of 97.18\%. The authors claim that their method is resilient to obfuscation and polymorphic code, but they admit that an adversary with knowledge of the model could deceive it by re-ordering the sections in a binary or ``adding vast amounts of redundant data". The method generated similar results to the then-current state-of-the-art with significantly fewer computational costs, as no disassembly or execution was required.

In later work \cite{Nataraj2011AAnalysis}, the same authors compared image texture analysis against dynamic analysis for malware classification. They found that on three independent datasets totalling 100,000 samples, texture analysis produced comparable classification performance while being 4000 times faster than dynamic analysis. The authors also suggest that image-based analysis works well despite obfuscation techniques, likely due to the use of fixed-size encryption keys that leave patterns or because malware families often implement their own obfuscation techniques, making their identification easier.

Gibert \cite{GibertConvolutionalClassification} trained convolutional neural networks with different architectures on a dataset of 10,868 malware samples with 9 classes (Microsoft Malware Classification dataset). Greyscale byteplots were created from the samples and then resized to 32×32 pixels before training the models. The author's motivation for using CNNs was to extract hierarchical image features as well as that the method employed by Nataraj et al. would not scale well with lots of data, while CNNs would. They achieved an accuracy of 99.76\% and showed that two convolutional layers with 64 and 128 kernels of size 3×3 was optimal. Their models, however, are highly vulnerable to the enlargement attacks discussed in this thesis due to the significant resizing operation.

\subsubsection{Adversarial Attacks}
A few other works have also focused on attacks against visualisation-based malware detection while ensuring that the adversarial malware samples are still executable and maintain functionality. Khormali et al. \cite{Khormali2019COPYCAT:Detection} proposed adversarial padding for untargeted misclassification attacks and sample injection for targeted misclassification. Adversarial padding concatenates an adversarial example generated by a traditional image-based adversarial generator to the end of the image, while sample injection concatenates samples from other classes or benign samples to the end of the image. All the images are resized to a constant, unspecified width and height. From testing their methods on two datasets, they achieved untargeted misclassification rates of 98.9\% for Windows binaries. The authors showed that traditional methods of generating adversarial examples for image classification models are not suitable for malware detection, as the adversarial examples are no longer executable as the binary code is corrupted. They also demonstrated that training the model on the adversarial examples negated their effect on classification performance.

Benkraouda et al. \cite{Benkraouda2021AttacksExecutability} claim that these types of adversarial samples that rely on end-padding can be easily detected and dealt with by cropping the sample during pre-processing. Instead of padding, they first generate an optimal adversarial example in the image space using traditional techniques, and then attempt to automatically produce a matching image by adding semantic NOP instructions throughout the malware sample. Semantic NOPs include neutral instructions such as adding zero to a register, jumping to the immediate next instruction, ANDing 1 with a register, or just normal NOP instructions. The adversarial examples took up to a few hours each to compute, but the authors were able to achieve a 98.9\% success rate. However, a very small dataset of 300 samples was used to evaluate the method.

\subsubsection{MIL}
Ilse et al. \cite{Ilse2018Attention-basedLearning} proposed an embedding-based MIL algorithm that uses an attention mechanism to aggregate instances together to form a bag representation. The attention mechanism provides a measure of how much an individual patch contributes to the bag representation through a weighted average, where the weights are learned by a two-layer neural network. The benefit of the attention mechanism is that it allows for interpretability and analysis of individual image patches while being a flexible permutation-invariant aggregator for embedding-based MIL.

\subsubsection{MIL Imaged-Based Malware Analysis}
To the extent of our knowledge, no research has formally investigated the use of multiple instance learning to classify malware images. However, there are two papers from the literature which informally use techniques similar to the MIL paradigm. 

A collaboration between Microsoft and Intel by Chen et al. \cite{Chen2020STAMINA:Classification} leveraged malware visualisation and deep transfer learning to efficiently classify malicious and benign samples at a large scale. They resized the malicious images to 224×224 or 299×299 and achieved an accuracy of 95.97\% on a dataset of 782,224 samples. The authors noted that the benign files were larger on average, and resizing large images would result in texture and information loss. They therefore chose to divide the benign images into patches, predict a score for each patch, and then aggregate the predictions by taking the mean. This process is essentially an instance-based MIL approach. Although the authors reported excellent performance, the instance-based approach is less powerful as the predictions rely solely on the information of single patches, which could be a problem if a feature is split across multiple patches.

He and Kim \cite{He2019MalwareTechniques} investigated using Spatial Pyramidal Pooling (SPP) to allow a CNN to classify malware byteplots of any size, recognising that any form of lossy transformation of the input image would make a technique prone to information loss that could be exploited by an attacker. SPP pools features extracted from pre-defined patch sizes to generate a fixed-length output for an arbitrary sized input. However, to deal with the memory constraints, the authors used a ``divide and conquer" approach. Similar to \cite{Chen2020STAMINA:Classification}, their approach was identical to instance-based MIL, where patches were classified individually as benign or malicious. However, in this case the authors used the standard MIL assumption so that a sample was malicious if any of the patches were classified as malicious. In contrast to \cite{Chen2020STAMINA:Classification}, the authors reported ``devastating failure” of this technique, attributed to malware features being split across patches and training the model on whole images while evaluating it on image patches. Their results indicate that for malware, individual image patches do not contain sufficient information to classify a whole sample. The authors recommend that a ``divide-and-merge" technique (such as embedded-based MIL) might be more appropriate.

%% file: assets/resizing_data_loss/sample_histogram.tex
\definecolor{myred}{HTML}{FF7F7F}
\definecolor{myblue}{HTML}{7F7FFF}

\begin{tikzpicture}[font=\footnotesize]
\begin{axis}[
    xlabel={Pixel Intensity},
    xlabel near ticks,
    ylabel near ticks,
    ylabel={Number of Pixels},
    ymin=0,
    ybar,
    enlarge x limits=0.01,
    legend entries={Original, Resized},
    legend pos=north west,
    bar width=1,
    xlabel style={yshift=-7pt} 
]
\addplot[red!60!white, fill=red!60!white, fill opacity=0.7, draw opacity=0.7] table {assets/resizing_data_loss/original_hist.txt};
\addplot[blue!60!white, fill=blue!60!white, fill opacity=0.7, draw opacity=0.7, bar shift=-0.05cm] table {assets/resizing_data_loss/upscaled_hist.txt};
\end{axis}
\end{tikzpicture}

%% file: assets/resizing_data_loss/adversarial_loss.tex
\begin{tikzpicture}[font=\footnotesize]
\begin{axis}[
    xlabel={Adversarial Padded Size (square)},
    ylabel near ticks,
    ylabel={MI \% of original malware},
    xmin=1091, xmax=10000,
    ymin=0,
    legend pos=north west,
    ymajorgrids=true,
    grid style=dashed,
    xtick distance=2000,
    enlarge x limits=0.05,
    scaled x ticks=false,
]

\addplot[
    color=blue,
    ]
    coordinates {
    (1091, 14.6)(1500,13.55)(2000,13.42)(3000,12.72)(4000,11.45)(5000,11.42)(6000,8.73)(7000,6.34)(8000,2.72)(9000, 4.24)(10000, 3.98)
    };
    
\end{axis}
\end{tikzpicture}

%% file: sections/3_methodology.tex
\begin{figure*}[!t]
	\centering
	\includegraphics[width=0.9\textwidth]{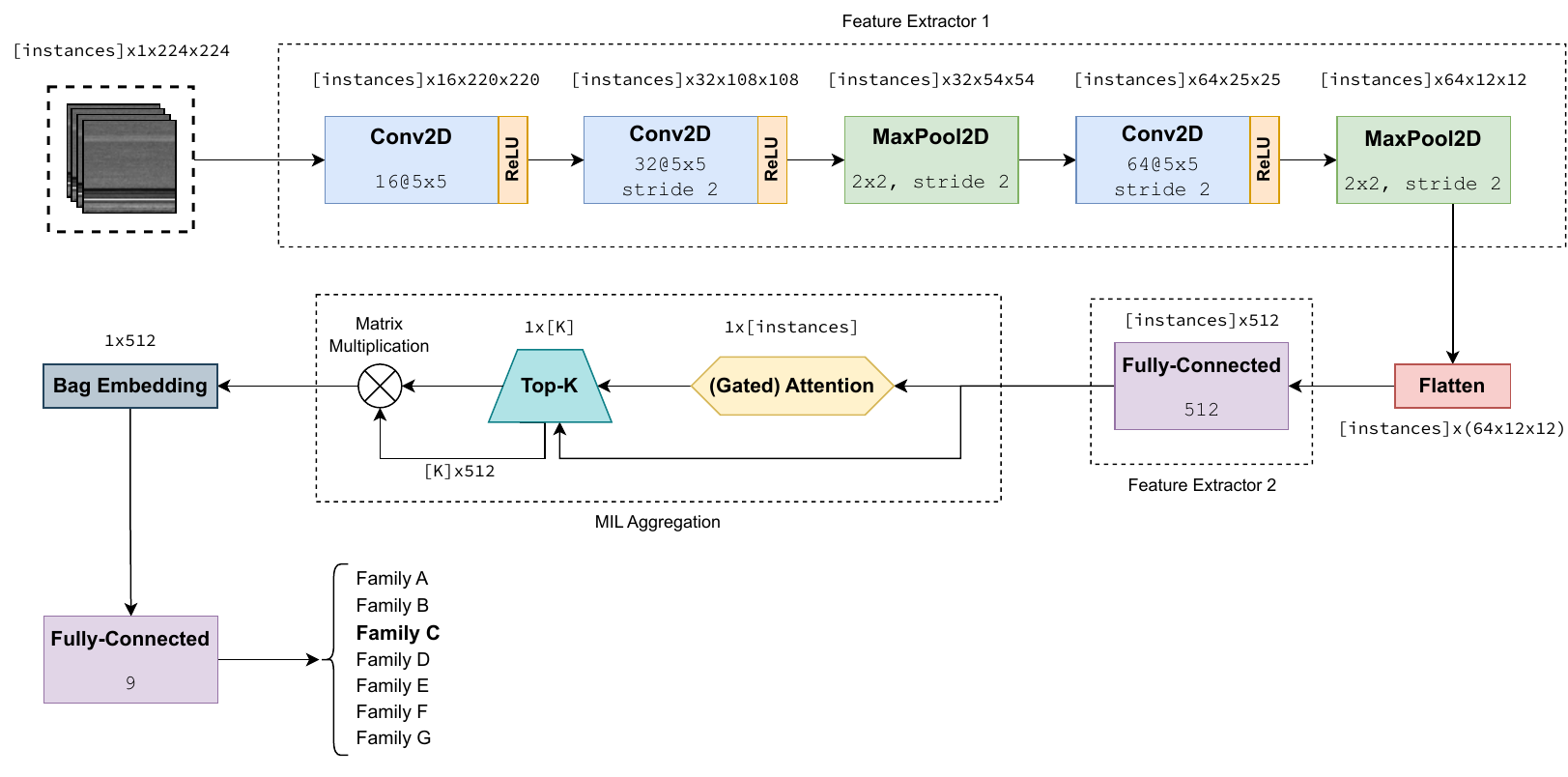}
	\caption{The model architecture for the MIL model.}
	\label{img:MIL_architecture}
\end{figure*}

To make a model that can classify malware of variable sizes in a memory efficient manner, we combine embedded-based multiple instance learning with a convolutional neural network. First, malware samples are converted into greyscale byteplots using the technique discussed in section \ref{subsec:Image-Based Malware Analysis}. Greyscale byteplots are used so that the model works on the raw malware bytes instead of a lossy image representation that would be vulnerable to adversarial enlargement. The images are then split into patches of 224×224 pixels to create bags of instances (see section \ref{subsec:MIL}). Bags are processed by the model one at a time, as they are often of different sizes. A CNN is used to extract a feature vector for each instance in a bag. Each feature vector is then assigned an importance score through the attention function, which is used to calculate a weighted average of the instances. Furthermore, to deal with adversarially enlarged images, only the instances with the top 12 attention scores are used in the averaging. This final bag representation is then classified using a fully-connected neural network. The model is illustrated in Figure \ref{img:MIL_architecture}.

Convolutional neural networks are used for their efficiency, translation equivariance, automatic hierarchical feature extraction, and proven performance in this task despite obfuscation techniques. Due to their translation equivariance and focus on local features, even if sections of the binary were moved around in an attempt at obfuscation, the model would still work well.

MIL is used to handle the large images of varying size due to its ability to effectively deal with non-i.i.d instances and to keep the focus of the model on local features despite adversarial enlargement attacks.

As the primary aim of this paper is to evaluate the effectiveness of multiple instance learning against adversarial enlargement of malware byteplots, less focus is given to optimising the specific underlying neural network parameters (number of layers, number of kernels, size of kernels, size of dense layers, etc.) and hyperparameters (batch size, learning rate, epochs). These values are instead borrowed from the literature or decided on through a few empirical observations.

\subsection{MIL}
\label{subsec:MIL}
The network architecture for the MIL model is outlined in Figure \ref{img:MIL_architecture}, and is adapted from the attention-based deep MIL model by Ilse et al. \cite{Ilse2018Attention-basedLearning}. The proposed solution does not use the standard MIL assumption that a single image patch can be classified as malware or as a specific family to determine the label of the whole sample. Instead, the "weighted collective assumption" \cite{Foulds2010} is used to create a feature vector that is representative of the most important and distinctive parts of the malware, regardless of its size. By combining attention-based pooling and top-k selection, the proposed MIL aggregation function is invariant to both the size of a bag and the order of instances, as required.

To create a bag, the malware sample is converted into a greyscale byteplot with a fixed width of 224 pixels. This image is padded with black pixels to the next multiple of $224*224$, and then split vertically into patches of 224×224 pixels. Compared to splitting a square image into patches, this method preserves the continuity of bytes and is applicable as the images are not natural. Each patch is then converted to a tensor and scaled to [0,1].

It is not directly possible to use mini-batching with the bags, as each bag might have a different number of instances, resulting in a non-constant input shape which cannot be placed inside a standard tensor for neural network computation. Therefore, a batch size of 1 is used, and the batch dimension is replaced with the instance dimension. Specifically, instead of the input shape (batch\_size, 224, 224, 1) the data loader loads (1, instances, 224, 224, 1). The first dimension is then dropped before processing the instances in parallel through the convolutional and pooling layers. The final result is a tensor of instance features of shape (instances, 512).

\subsubsection{Attention}
The attention mechanism from Ilse et al. \cite{Ilse2018Attention-basedLearning} allows the model to determine which image patches carry more importance for classifying the malware family. Compared to normal attention, it is assumed that instances are independent and not sequentially dependent. Attention is calculated using two fully-connected layers and a $tanh()$ activation function. A softmax function is applied over the attention vectors so that they lie in the range [0,1] and sum to one, making them invariant to the size of the bag. The effect of a gated attention mechanism to better suppress irrelevant instances is also tested. 

\subsubsection{Sub-batching}
Using a batch size of 1 means the GPU memory requirements are based on the size of a single byteplot. However, as discussed previously in Section \ref{subsec:Problems with Resizing}, a high amount of memory would be required to store the intermediate feature maps for large malware, as splitting into patches does not change the number of pixels. Therefore, sub-batching is implemented by splitting each bag into groups of 60 instances and processing the sub-batches through the convolutional and pooling layers one at a time. The memory requirements are therefore reduced to 60 images of shape (224,224,1).

\subsubsection{Gradient Accumulation}
Using a batch size of 1 in a multi-class classification setting can lead to ineffective gradient and weight updates, as the model only sees a sample from one class every optimisation step. However, due to memory limitations, a higher batch size cannot be used. Gradient accumulation is used instead, which calculates and collects the gradients after every bag is processed during training, but only updates the model parameters after a certain number of bags - effectively simulating a higher batch size. The number of bags is equivalent to a batch size, and was set to 48 in the experiments.

\subsubsection{Top-k Selection}
The model was originally trained without top-k selection, which resulted in very poor testing performance (3-5\% accuracy). After analysing the attention weights in Figure \ref{fig:attention_weights_}, we noticed that even though each patch of noise was given a low attention score of 0.0003, there were so many of them in an adversarially enlarged sample that they added up to over 80\% of the attention in some cases. This was likely causing the poor performance, so a top-k selection was implemented to only use the instances with the top 12 attention scores to create the final bag representation. This number was chosen as it is the average number of instances per bag in the Microsoft Malware Classification dataset. Top-k selection makes the MIL aggregation function robust to adversarial enlargement, which in turn makes the whole model robust.

\begin{figure*}[!t]
	\centering
	\includegraphics[width=0.8\textwidth]{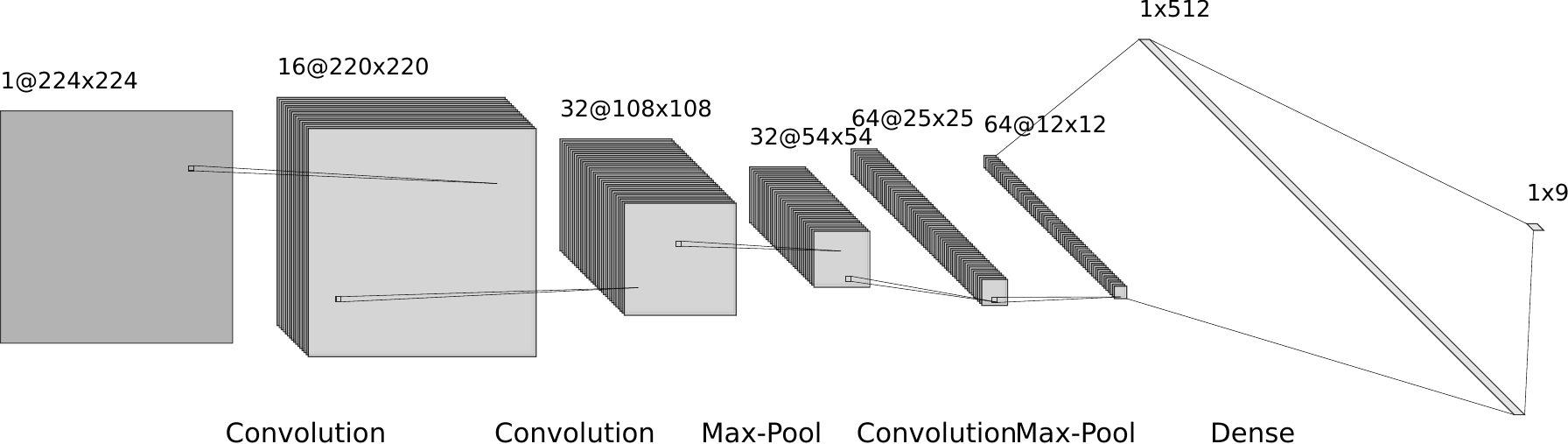}
	\caption{Base model CNN architecture.}
	\label{img:base-cnn-architecture}
\end{figure*}

\subsection{Base}
To compare the MIL model to the standard use of CNNs from the literature and to empirically test the effect of adversarial enlargement on these methods, a baseline model that uses image resizing is also implemented. The network architecture for the Base model is outlined in Figure \ref{img:base-cnn-architecture}.

Greyscale malware byteplots were created in the standard way, with the width and height set to the square root of the length of the malware. The square root was chosen for simplicity to create square images that would not be distorted too much during resizing. Additionally, using the square root has been shown to perform similarly to other image width choices \cite{Chaganti2022Image-basedClassification}. The images were then resized to 224×224 using bilinear interpolation, converted to a tensor, and scaled to [0,1]. 

\subsection{Adversarial Enlargement}

\subsubsection{Threat Model}
The aim of adversarial machine learning is to modify an input $x$ in such a way that it is misclassified by the model. I assume that the goal of the attacker is to conduct untargeted misclassification attacks at test time. Untargeted misclassification is defined as generating adversarial examples that force the output of the model to be any class other than the correct one for a given sample \cite{Khormali2019COPYCAT:Detection}. Furthermore, I assume that the attacker has partial knowledge of the model, limited to the fact that the model is CNN-based MIL. The attacker lacks full access to the training data, exact parameters, and gradients, making it a grey-box attack. The capabilities of the attacker include changing the structure of a malware sample or changing its contents by adding large amounts of non-functional data in any location. Of course, the adversary could create a surrogate model trained with similar data to perform more advanced single-step adversarial attacks, but these have been shown to be non-transferrable between models \cite{Suciu2018ExploringDetection}.

\subsubsection{Generating Examples}
\begin{figure*}[!t]
    \centerline{
    \subfloat[\centering Original Malware]{{\includegraphics[width=0.3\textwidth]{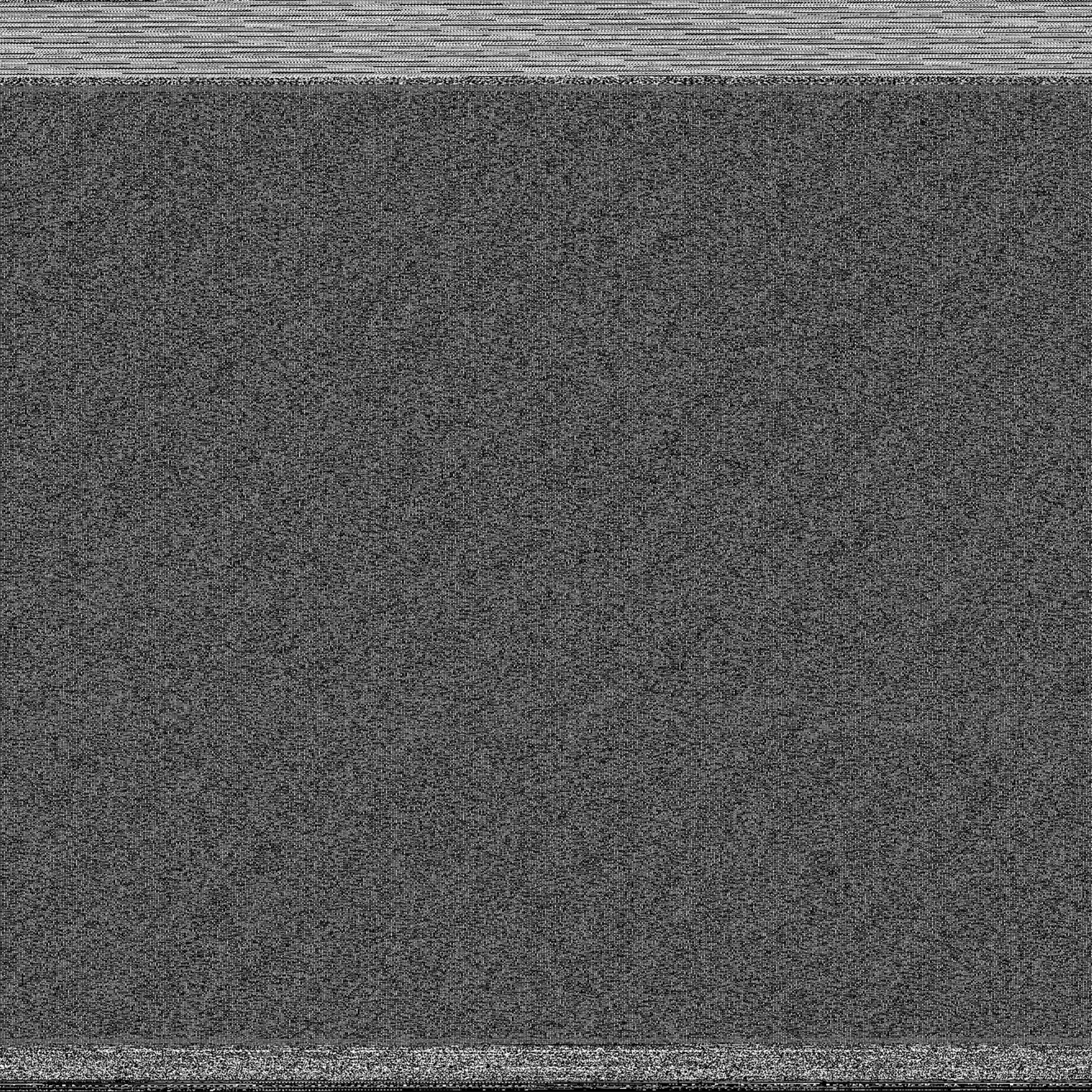} }}%
    \subfloat[\centering Zero-padding enlargement]{{\includegraphics[width=0.3\textwidth]{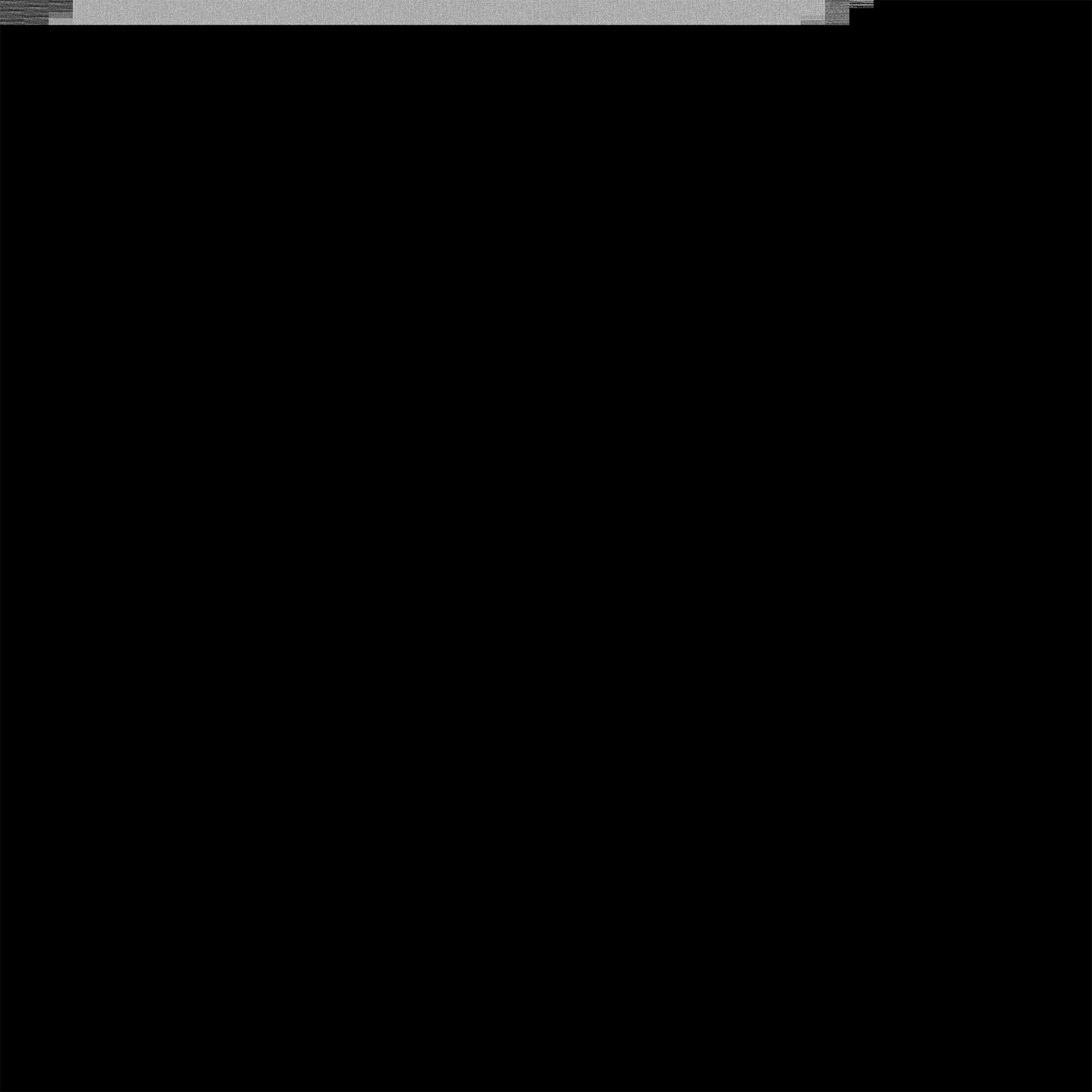} }}%
    \subfloat[\centering Uniform noise padding enlargement]{{\includegraphics[width=0.3\textwidth]{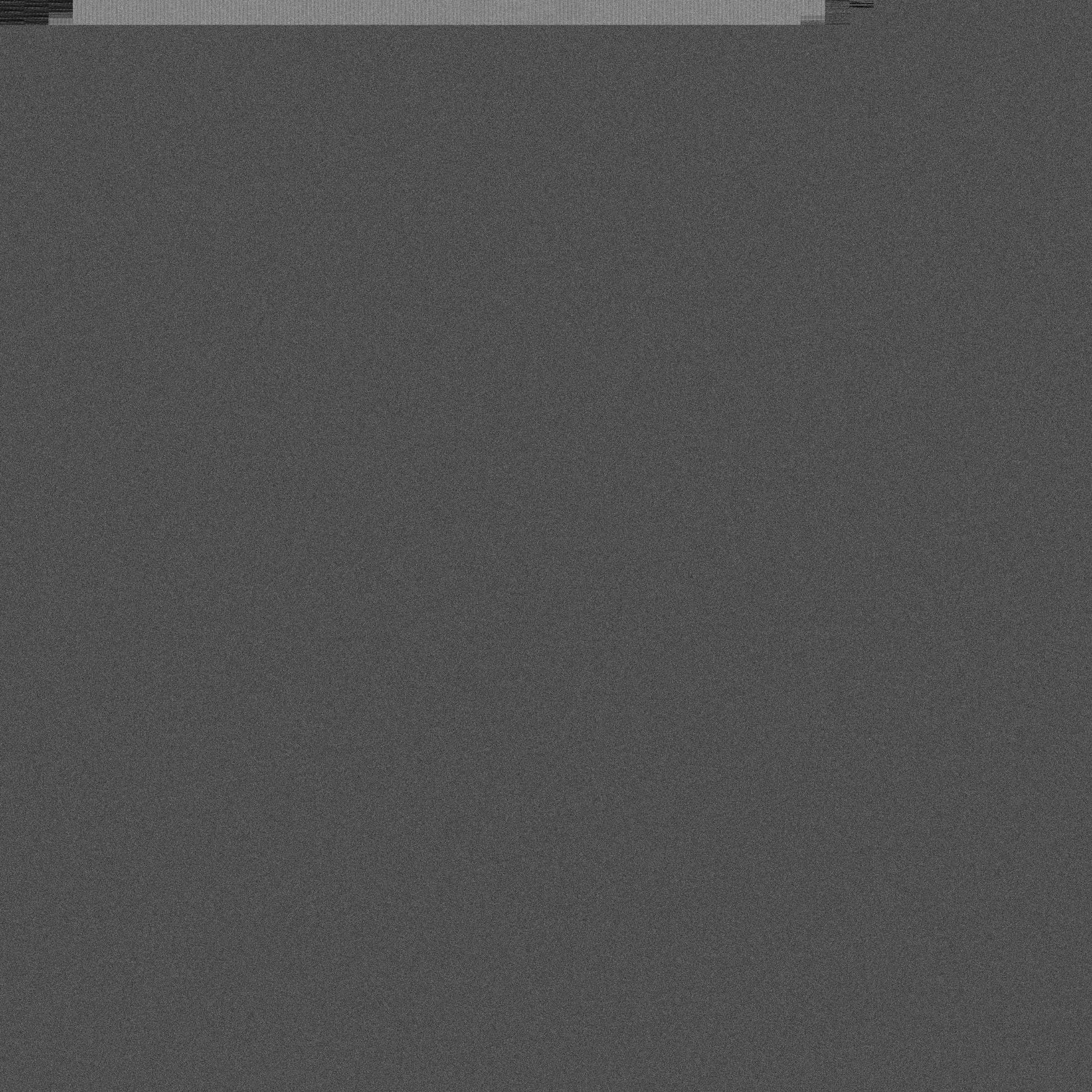} }}%
    }
    \caption{A malware sample enlarged to 10000×10000px with black pixels and noise, visualised as squares of size 224×224px (not to scale) to see the effect more clearly.}%
    \label{fig:adversarial_examples}%
\end{figure*}

As a proof-of-concept to generate adversarial enlargement examples, redundant data is appended to the end of the malware samples to enlarge them to 10000×10000px. Appending data was chosen for ease of implementation and is the simplest method that does not violate the syntactic and semantic constraints of an executable. The malware will still be executable and maintain the same functionality. The impact of appending black pixels as well as uniform noise is tested to see which has a greater effect on classification (examples in Figure \ref{fig:adversarial_examples}).

Although adversarial examples were only generated by appending data (which could be considered trivial to overcome by truncating the sample), the proposed solution should be able to deal with other, more advanced methods of adversarial enlargement due to translation equivariance. As the MIL aggregation function is permutation-invariant and the CNN supports local translation equivariance, the location of where the redundant data has been added to the sample should not matter.

\subsection{Dataset}
The Microsoft Malware Classification dataset (BIG 2015) \cite{Ronen2018MicrosoftChallenge} was used to evaluate the proposed method. The dataset contains 10,860 known malware samples from 9 different families. The raw files are made available in hexadecimal format with the PE header removed for sterility. The hex data is converted into binary files before training. Additionally, many files in the dataset contain ``??" characters, which are removed. Some files are solely made up of question marks, and are therefore fully removed from the dataset.

\subsection{Experimental Setup}
All the models were developed and implemented using PyTorch 2.0.1 in Python 3.9. The models were trained on the University of Southampton Alpha cluster, which provides access to an Intel Xeon\textsuperscript{\tiny\textregistered} Silver 4216 CPU, 376GB DDR4 RAM, and an Nvidia Quadro RTX8000 GPU with 48GB of VRAM. The NumPy and pandas libraries were extensively used for data manipulation, TorchMetrics \cite{TorchMetrics2022} was used to calculate metrics, and the CometML \cite{CometML2023} platform was used to log results.

The dataset was split into stratified train and test sets with a $\sim$90:10 split (1000 test samples), which were kept separate to evaluate the performance of the models on unseen malware. For all models, cross-entropy loss and the Adam optimiser are used. Each model was trained for 20 epochs. All the models were trained on the unmodified, non-adversarial samples.

%% file: sections/4_results.tex
\begin{table*}[!t]
\renewcommand{\arraystretch}{1.3}
\caption{The results of the models evaluated on the Microsoft Malware Classification dataset as well as results from the literature. Note that \cite{Nataraj2011MalwareImages} was trained on a different dataset (MalImg). The F1-score is macro-averaged. The average inference speed is the network latency, and in the brackets is the total end-to-end time including overheads such as image resizing, bag creation, and adversarial enlargement.}
\label{table:results_table}
\centering
\begin{tabular}{cccccccccc}
\hline
 \makecell{Adversarial\\Enlargement} & Model & Train Duration & Accuracy & F1 & Loss & AUROC & \makecell{Avg Inference\\(w/ overheads) (ms)} & GPU Memory (GB) \\
\hline
    \multirow[c]{7}{*}[0in]{None}
    & Baseline & 0:07:48 & 0.963 & 0.924 & 0.233 & 0.992 & 0.112 (2.60) & 2.24 \\
    & MIL Attention & 0:42:30 & 0.973 & 0.932 & 0.200 & 0.993 &  2.45 (4.67) & 4.50 \\
    & MIL Gated Attention & 0:47:08 & 0.950 & 0.866 & 0.288 & 0.994 & 2.49 (4.65) & 4.50 \\
    \cline{2-9}
    & Nataraj et al. \cite{Nataraj2011MalwareImages} & - & 0.972 &  - &  - &  - &  1400 & - \\
    & Gibert \cite{GibertConvolutionalClassification} & - & 0.998 &  - &  - &  - &  - & - \\
    & Conti et al. \cite{Conti2022AVisualization} & - & 0.948 &  - &  - &  - &  - & - \\
    & Ni et al. \cite{Ni2018MalwareLearning} & - & 0.989 &  - &  - &  - &  - & - \\
    
    \hline
    
    \multirow[c]{3}{*}[0in]{Zeros}
    & Baseline & 0:05:56 & 0.228 & 0.041 & 11.279 & 0.424 &  0.125 (55.4) & 2.24 \\
    & MIL Attention & 2:06:51 & 0.940 & 0.860 & 0.397 & 0.988 & 186 (309) & 3.67 \\
    & MIL Gated Attention & 2:43:54 & 0.966 & 0.925 & 0.312 & 0.990 & 186 (300) & 3.66 \\

    \hline 

    \multirow[c]{3}{*}[0in]{Noise}
    & Baseline & 0:05:49 & 0.043 & 0.020 & 5.115 & 0.474 & 0.126 (360) & 2.24 \\
    & MIL Attention & 3:02:33 & 0.845 & 0.705 & 0.752 & 0.976 & 186 (378) & 3.67 \\
    & MIL Gated Attention & 4:25:33 & 0.827 & 0.665 & 0.738 & 0.962 & 187 (369) & 3.66 \\
\hline
\end{tabular}
\end{table*}

\subsection{Non-Adversarial}
The results from training and evaluating the models on the unmodified malware samples are presented in Table \ref{table:results_table}. 
\subsubsection{Baseline}
The results show that the baseline model performs similarly to the models from the literature, with an F1 score of 0.924 and only 2.24GB of GPU memory usage, making it a relatively light-weight and accurate classifier. This indicates that the implementation and methodology for these models allows for a fair comparison with the multiple instance learning models. Furthermore, it confirms that visualisation-based malware detection can work well.

\begin{figure*}[!t]
        \centering
        \subfloat[Baseline]{\includegraphics[width=0.32\textwidth]{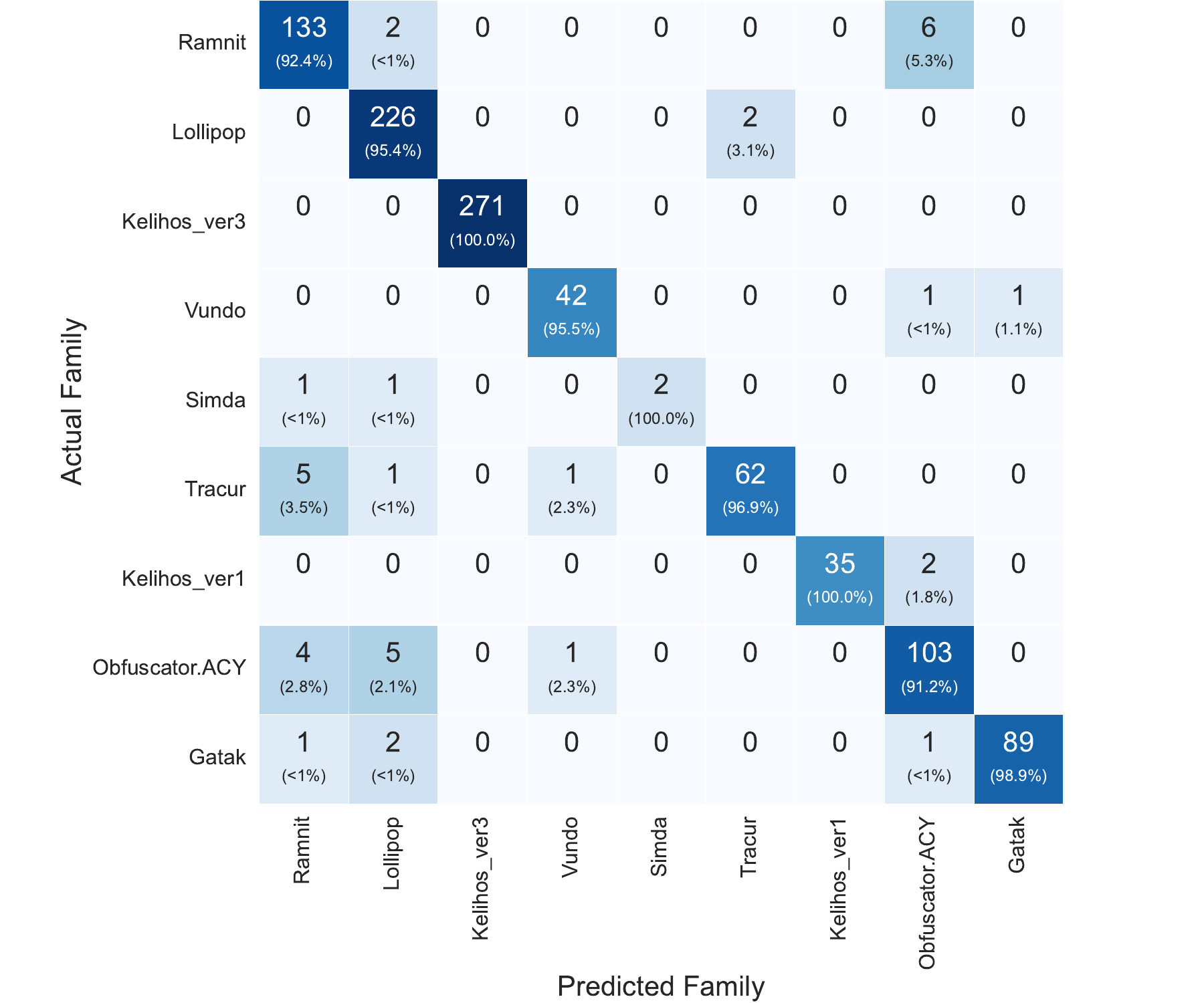} \label{fig:CM_baseline_normal}}%
        \hfill
        \subfloat[Attention]{\includegraphics[width=0.32\textwidth]{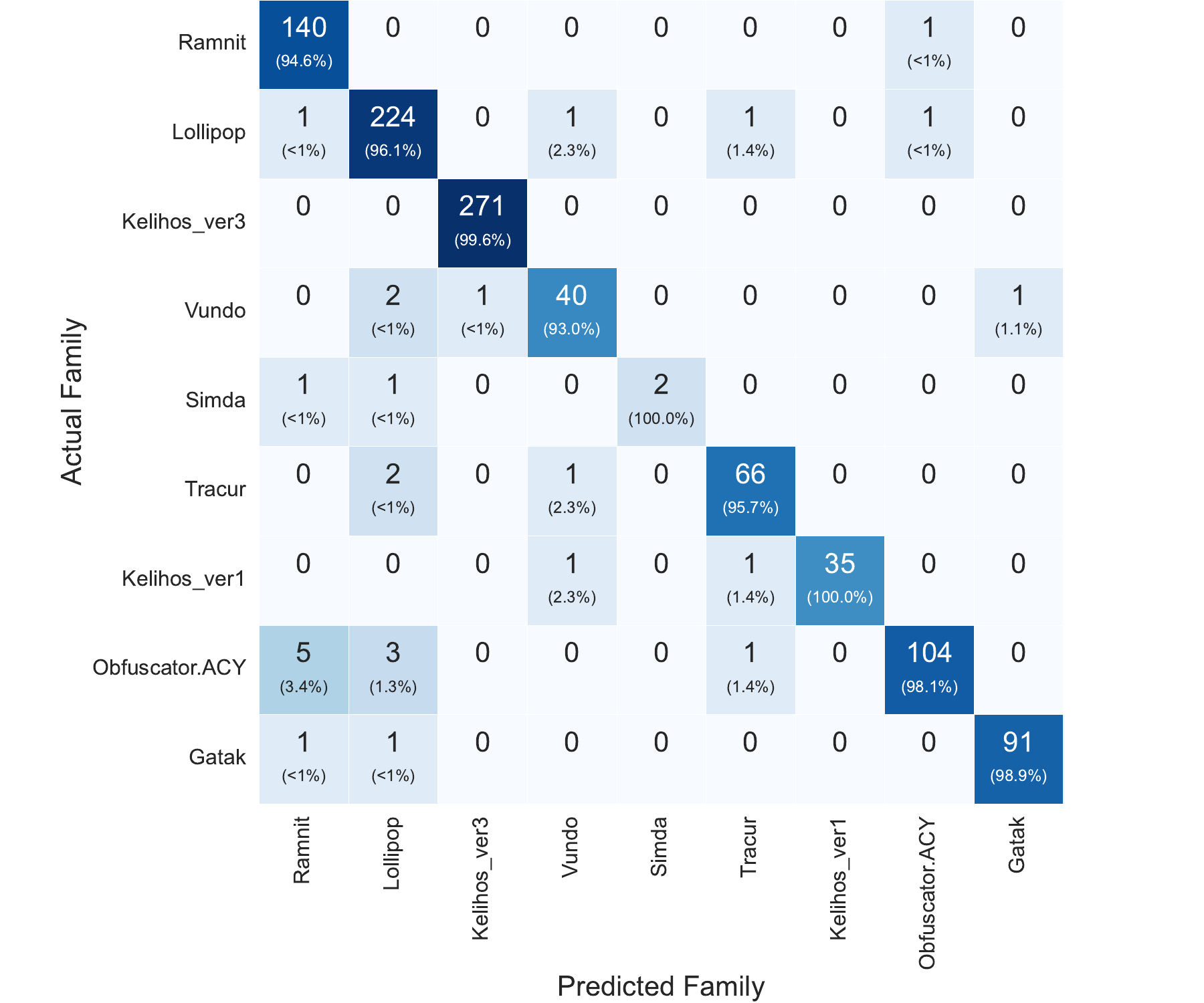} \label{fig:CM_attention_normal}}%
        \hfill
        \subfloat[Gated Attention]{\includegraphics[width=0.32\textwidth]{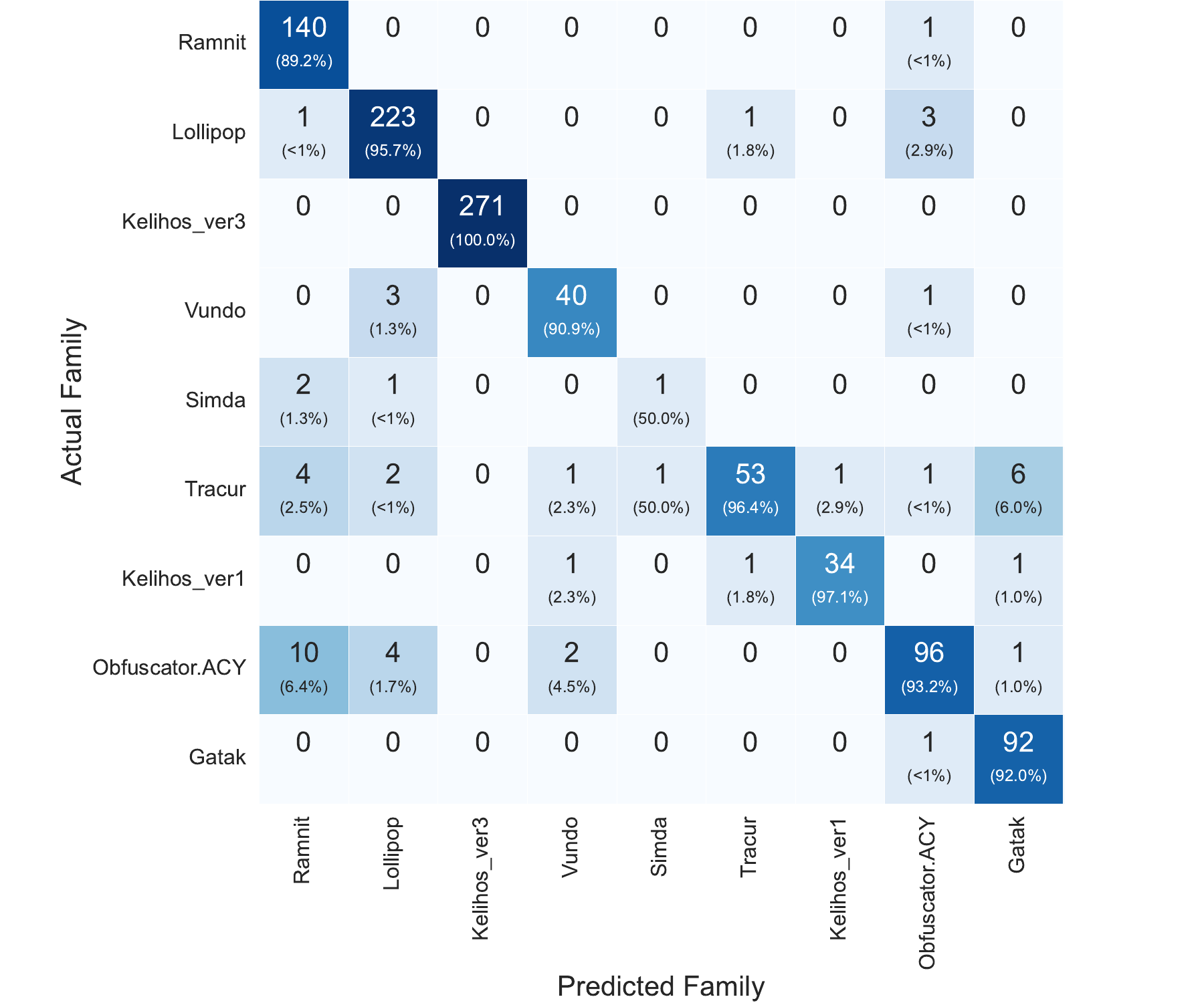} \label{fig:CM_Gatedattention_normal}}%
        
	\caption{Confusion matrices for the models evaluated on unmodified, non-adversarial samples. Percentages are calculated by column, indicating the precision per class.}%
    \label{fig:CM_normal}%
\end{figure*}

From the confusion matrix in Figure \ref{fig:CM_baseline_normal}, it can be seen that the baseline CNN model struggled slightly with confusing Obfuscator.ACY samples for Ramnit and Lollipop. However, overall there are few off-diagonal values.

\subsubsection{MIL}
The multiple instance learning model using the attention mechanism performed better than the model with the gated attention mechanism, with F1 scores of 0.932 and 0.866, respectively. Both models had similar costs, taking almost the same amount of time to train and the exact same amount of GPU memory. 

Compared to the baseline model, the attention MIL model performed slightly better across the reported metrics. Except for AUROC, the gated attention MIL model performed the worst. The models show good performance for classifying standard, non-adversarial malware. By optimising the model architecture and hyperparameters more, it is highly likely that an MIL model could achieve even better classification performance.

The main cost of the MIL models is the significant increase in the time required to train the model. However, the training duration includes time taken to load the datasets, which could take around 10 minutes for the MIL model as each sample was loaded into memory and processed into a bag. Considering the average bag size of the training dataset ($\sim$12.6 instances), this equates to a sub-linear increase in time compared to the baseline CNN model. The GPU memory requirements did not increase as much, however, demonstrating the ability of MIL to deal with large images in a memory efficient manner.

The confusion matrices in Figure \ref{fig:CM_normal} show that the attention MIL model didn't struggle with any class specifically, but suffered from more general classification issues, which could be resolved with hyperparameter tuning, for example. The gated attention model performed slightly worse overall, with more distributed misclassifications for the Tracur family. This could be due to the gated attention mechanism overfitting to the data more, as gated attention has a higher complexity than normal attention.

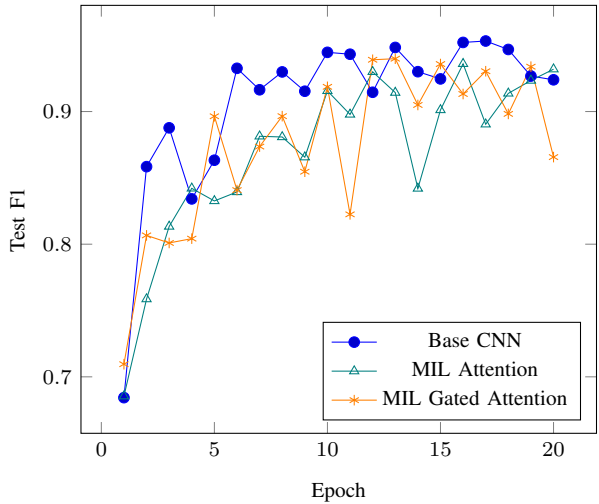
\begin{figure}[!t]
	\centering
	\input{assets/results/Normal_F1_graph}
	\caption{Test F1 scores across the epochs for the models trained on the unmodified samples.}
	\label{figure:normal_f1_epoch_graph}
\end{figure}

The F1 score trends in Figure \ref{figure:normal_f1_epoch_graph} show that the base model is more stable with higher F1 scores during training, while the MIL models vary in performance. This is probably due to the increased amount of noise for MIL, as the size of the input increased significantly while the model complexity remained relatively similar. The base CNN and MIL models show rapid improvement in the first few epochs, and then level off. However, the MIL models still show gradual improvement over the epochs, suggesting a potential for better results over additional epochs.

\subsection{Adversarial}
The results from training and evaluating the models on the adversarially enlarged malware samples are presented in Table \ref{table:results_table}. 

\begin{figure*}[!t]
        \centering
        \subfloat[Baseline]{\includegraphics[width=0.32\textwidth]{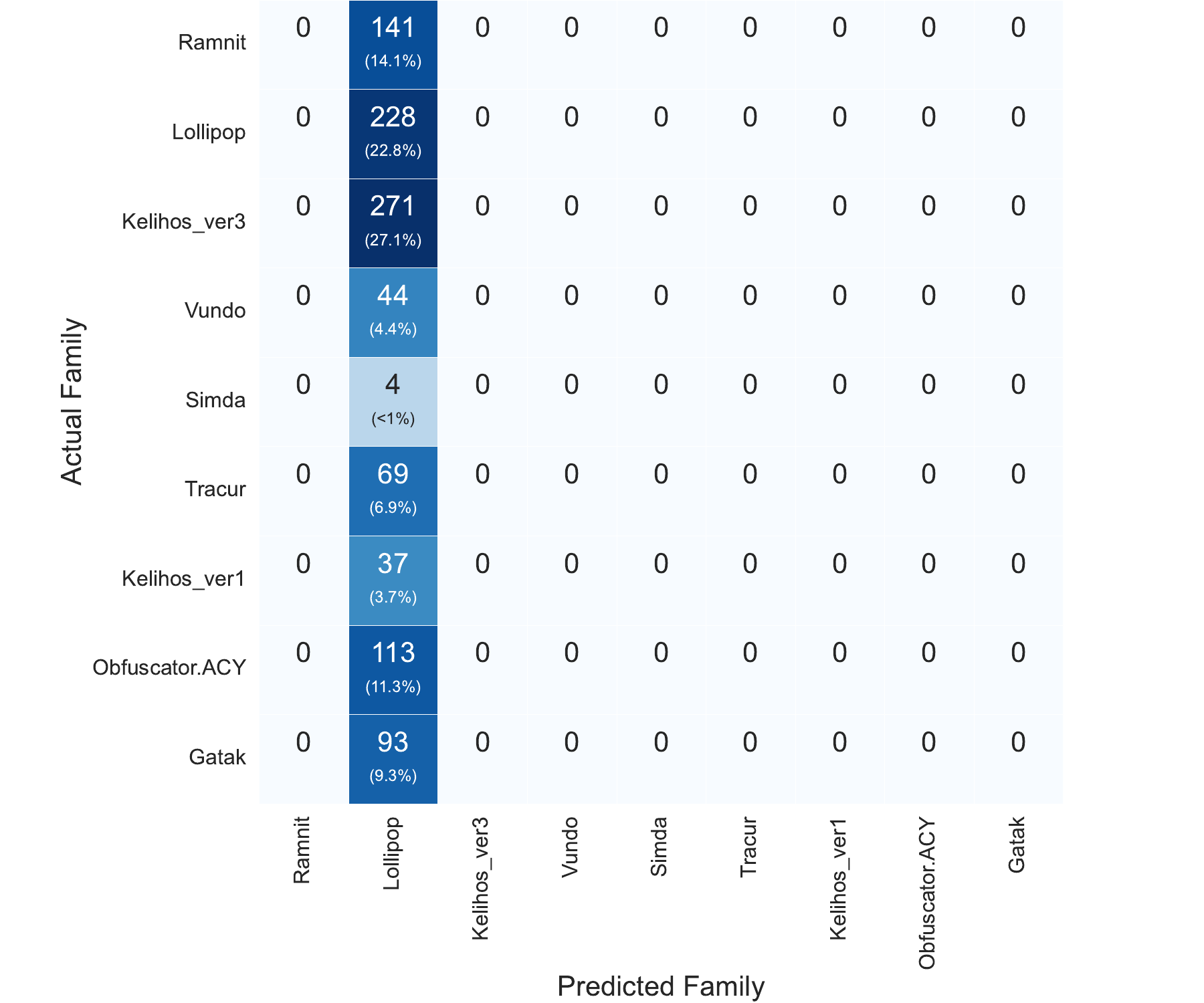} \label{fig:CM_baseline_zero}}%
        \hfill
        \subfloat[Attention]{\includegraphics[width=0.32\textwidth]{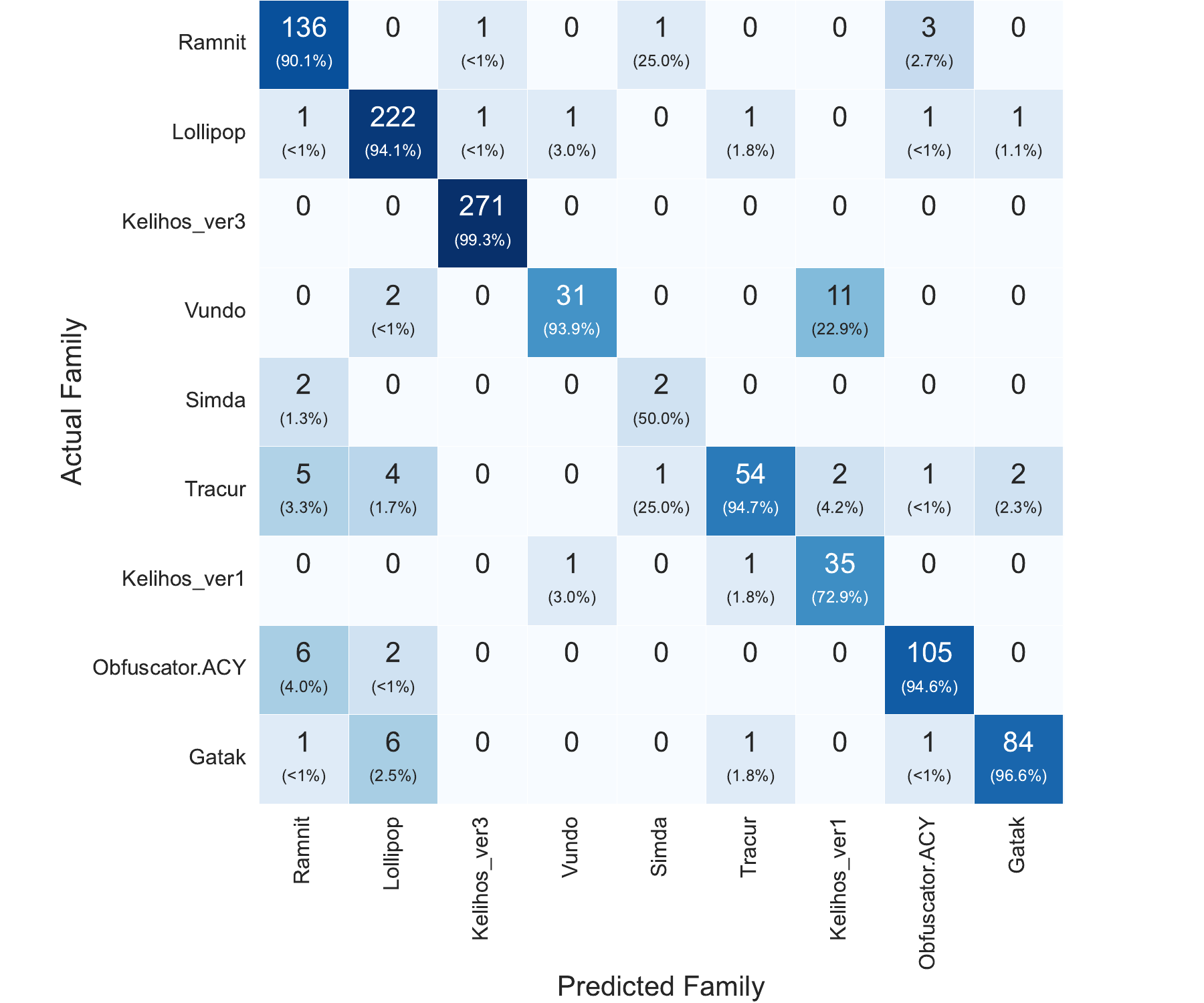} \label{fig:CM_attention_zero}}%
        \hfill
        \subfloat[Gated Attention]{\includegraphics[width=0.32\textwidth]{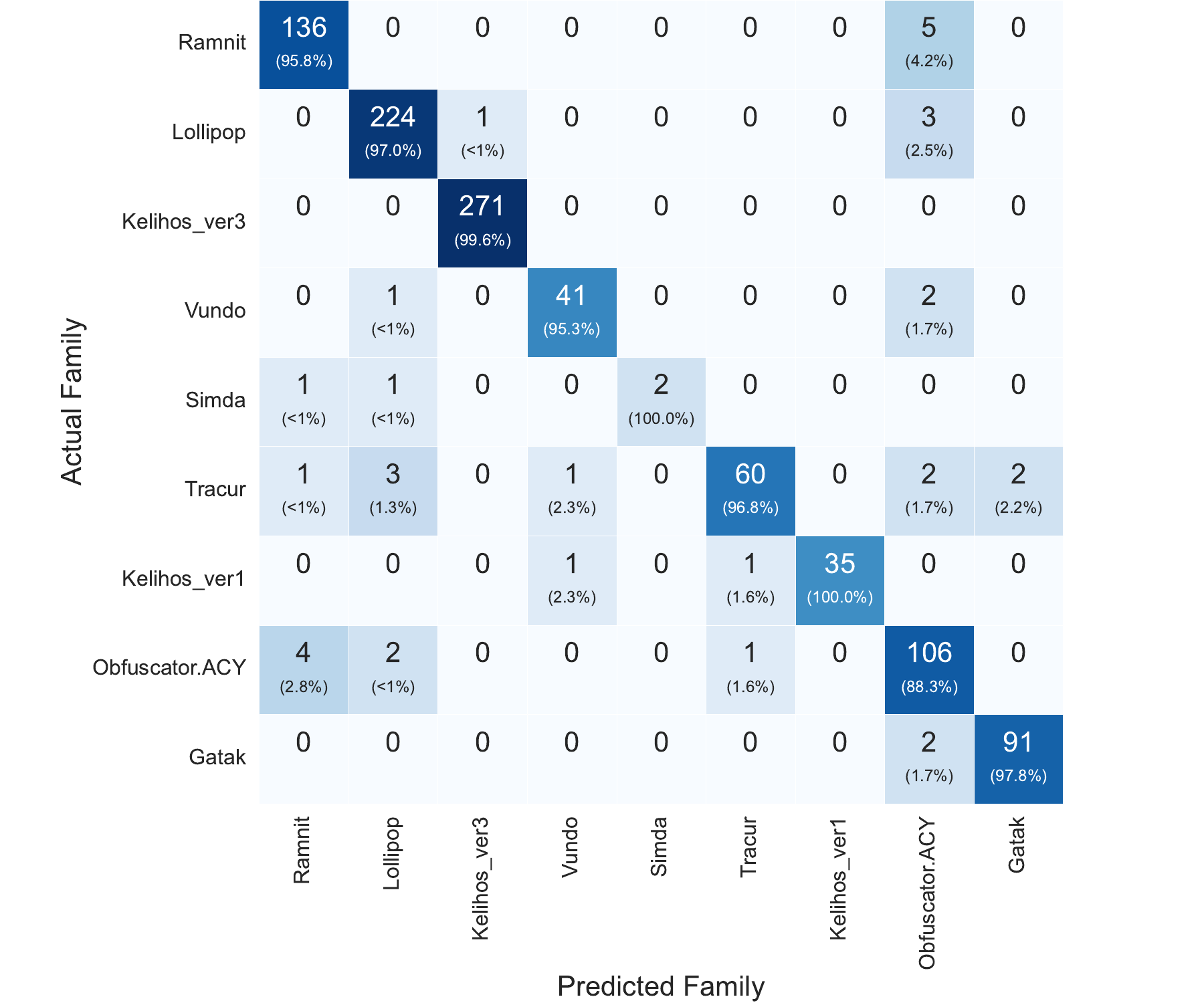} \label{fig:CM_Gatedattention_zero}}%
        
	\caption{Confusion matrices for the models evaluated on the samples adversarially enlarged with zeroes. Percentages are calculated by column, indicating the precision per class.}%
    \label{fig:CM_zero}%
\end{figure*}

\begin{figure*}[!t]
        \centering
        \subfloat[Baseline]{\includegraphics[width=0.32\textwidth]{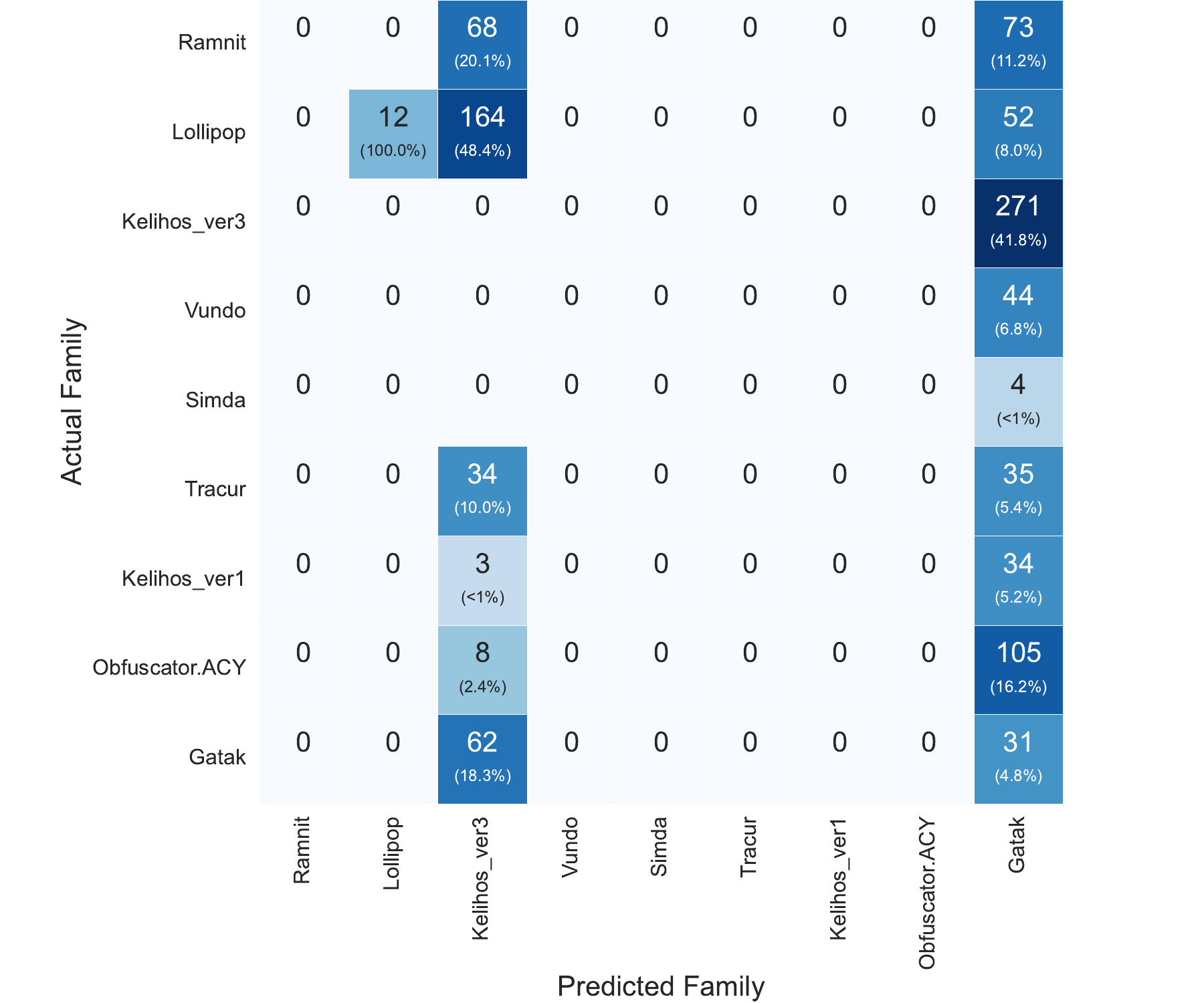} \label{fig:CM_baseline_noise}}%
        \hfill
        \subfloat[Attention]{\includegraphics[width=0.32\textwidth]{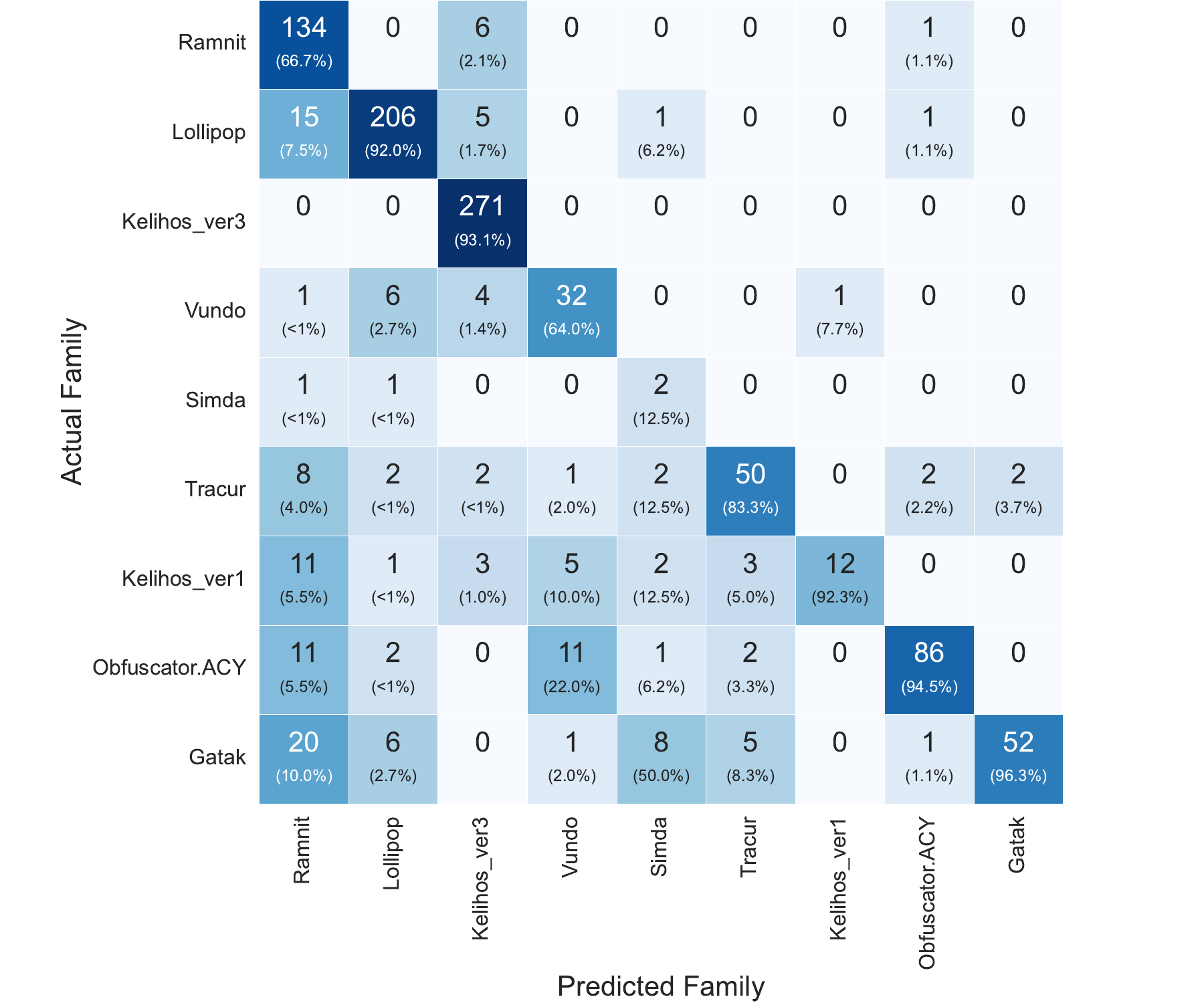} \label{fig:CM_attention_noise}}%
        \hfill
        \subfloat[Gated Attention]{\includegraphics[width=0.32\textwidth]{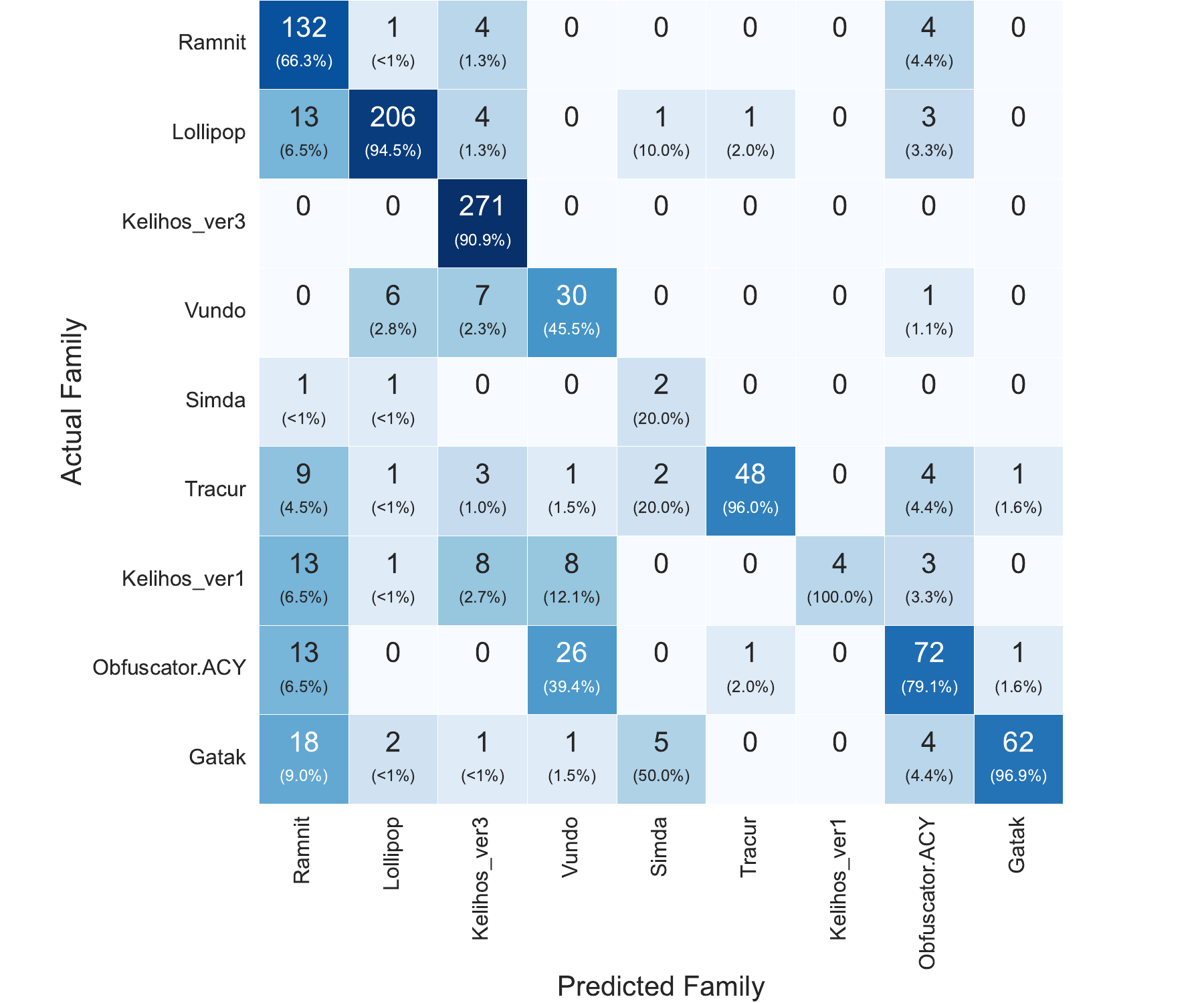} \label{fig:CM_Gatedattention_noise}}%
        
	\caption{Confusion matrices for the models evaluated on the samples adversarially enlarged with uniform noise. Percentages are calculated by column, indicating the precision per class.}%
    \label{fig:CM_noise}%
\end{figure*}

\subsubsection{Baseline}
The results for the baseline classifier tested on the adversarial malware samples show the catastrophic failure of the baseline model to accurately classify the byteplots. The CNN model achieved an F1 score of 0.041 for classifying samples enlarged with null-bytes, and an F1 score of 0.020 for samples enlarged with uniform noise. The model performed equally poorly for zero and noise-enlarged samples, with noise-padding having a slightly greater effect on classification accuracy. 

These results empirically demonstrate that lossy malware visualisation techniques such as resizing are not appropriate for malware classification, as they are vulnerable to untargeted misclassification attacks via the adversarial enlargement of samples. The crucial loss of data impacts the classifier's ability to correctly predict malware families.

Based on the AUROC values of 0.424 and 0.474, the adversarially-enlarged samples actively make the base CNN model worse at discriminating between malware families than a completely random classifier. This is likely because it interferes more with the ability of the kernels to detect small, specific features in the input. 

The confusion matrices in Figure \ref{fig:CM_baseline_zero} and Figure \ref{fig:CM_baseline_noise} illustrate the reason for the low accuracy and F1 metrics. The model predicts all the null-byte enlarged malware samples as a single class, Lollipop, while splitting the noise-enlarged sample predictions across just three classes; Gatak, Kelihos\_ver3, and Lollipop. The strategy of classifying every sample as Lollipop indicates that there is not enough information left in the resized samples for the model to identify discriminatory features. The uniform noise likely causes predictions to fall into Gatak and Kelihos\_ver3 because their samples typically seem to contain noisy features.

\subsubsection{MIL}
The MIL models trained on unmodified samples are still able to classify some adversarially enlarged samples relatively well, indicating the degree of effectiveness of the solution. In contrast to the non-adversarial setting, the gated attention mechanism performed the best for null-byte enlargement, with an F1 score of 0.925. However, for noise-enlargement the non-gated attention mechanism performed best again with an F1 score of 0.705. Although the F1-score for noise-enlarged test case is relatively low, it is significantly better than the baseline model. Additionally, the AUROC metrics are still high, indicating good discriminatory power.

Malware samples adversarially enlarged with uniform noise generally decreased the performance of the models more than samples enlarged with null bytes. Furthermore, while the models were generally good at classifying the obfuscated class (Obfuscator.ACY), the addition of noise to the samples caused relatively high misclassification rates for these samples.

In terms of cost, the GPU memory requirements have only slightly increased compared to the non-adversarial MIL models while the average bag size has increased from 12.6 to 2000. This again shows the ability of the proposed solution to process large images in a memory-efficient manner. However, the amount of time required to train the models has increased considerably to hours instead of minutes for MIL models.

The confusion matrices in Figure \ref{fig:CM_zero} and Figure \ref{fig:CM_noise} clearly show the models struggle with adversarial noise and generalise the features from the Ramnit family too much. This could be due to the gated attention sometimes suppressing parts of the malware instead of the adversarial noise.

\begin{figure}[!t]
	\centering
	\input{assets/results/Adv_F1_graph}
	\caption{Test F1 scores across the epochs for the models trained on the adversarial samples.}
	\label{figure:adv_f1_epoch_graph}
\end{figure}
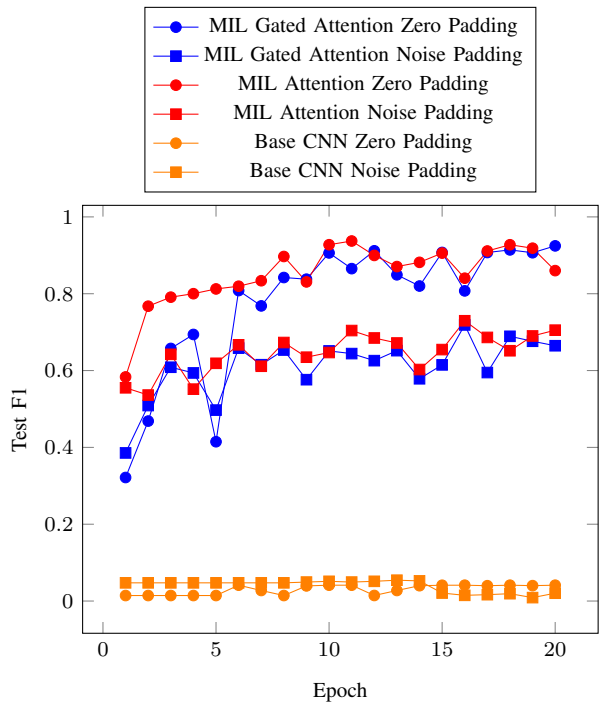

The F1 scores displayed in Figure \ref{figure:adv_f1_epoch_graph} show no significant improvement in the ability of the baseline models to classify the adversarial examples, while the MIL models show a fluctuating but increasing performance over the epochs. 

\begin{figure*}[!t]
        \centering
        \subfloat[Attention]{\includegraphics[width=0.45\textwidth]{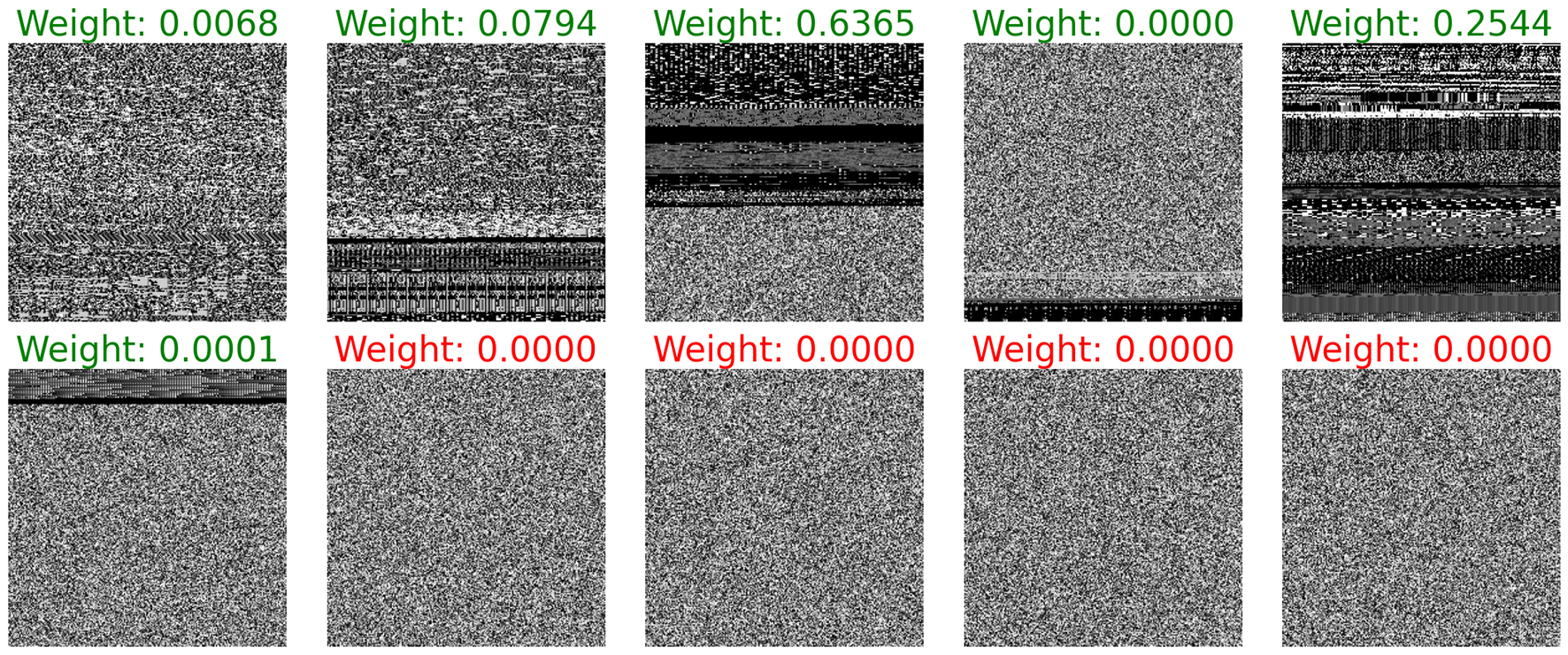} \label{fig:attention_weights_}}%
        \hfill
        \subfloat[Gated Attention]{\includegraphics[width=0.45\textwidth]{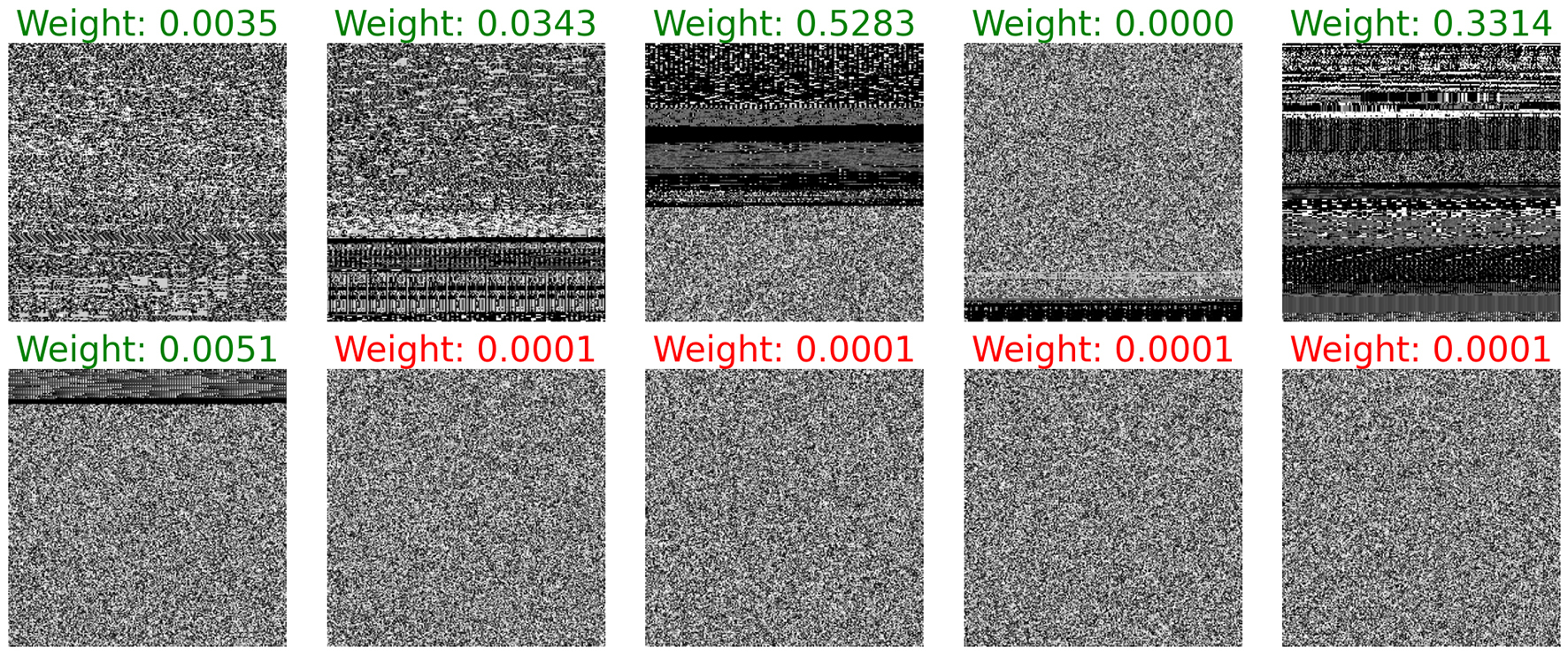} \label{fig:gated_attention_weights_}}%
        
	\caption{Gated and normal attention weights for the first 10 patches of an adversarially enlarged malware sample. The first six patches with green text are the original malware, and the red patches are noise padding.}
	\label{img:attention-weights-comparison}
\end{figure*}

By comparing the attention weights from the attention and gated attention models for the same sample in Figure \ref{img:attention-weights-comparison}, it is clear that the normal attention mechanism attributes a lower weight to the adversarial noise than the gated attention mechanism does (0.0 vs 0.0001 and 0.6365 vs 0.5283). This can be attributed to the gating mechanism overfitting.

%% file: assets/results/Normal_F1_graph.tex
\begin{tikzpicture}[font=\footnotesize]
\begin{axis}[
    xlabel={Epoch},
    ylabel near ticks,
    ylabel={Test F1},
    legend pos=south east,
    legend style={font=\footnotesize},
]

\addplot table [x=Epoch, y=Base CNN, col sep=comma] {assets/results/f1_scores_non_adversarial.csv};
\addlegendentry{Base CNN};

\addplot[color=teal, mark=triangle] table [x=Epoch, y=MIL Attention, col sep=comma] {assets/results/f1_scores_non_adversarial.csv};
\addlegendentry{MIL Attention};

\addplot[color=orange, mark=asterisk] table [x=Epoch, y=MIL Gated Attention, col sep=comma] {assets/results/f1_scores_non_adversarial.csv};
\addlegendentry{MIL Gated Attention};

\end{axis}
\end{tikzpicture}

%% file: assets/results/Adv_F1_graph.tex
\begin{tikzpicture}[font=\footnotesize]

\pgfplotsset{every axis legend/.append style={at={(0.5,1.03)}, anchor=south}}

\begin{axis}[
    xlabel={Epoch},
    ylabel near ticks,
    ylabel={Test F1},
    legend style={font=\footnotesize},
]

\addplot[color=blue, mark=*] table [x=Epoch, y=MIL Gated Attention Zero Padding, col sep=comma] {assets/results/f1_scores_adversarial.csv};
\addlegendentry{MIL Gated Attention Zero Padding};

\addplot[color=blue, mark=square*] table [x=Epoch, y=MIL Gated Attention Noise Padding, col sep=comma] {assets/results/f1_scores_adversarial.csv};
\addlegendentry{MIL Gated Attention Noise Padding};

\addplot[color=red, mark=*] table [x=Epoch, y=MIL Attention Zero Padding, col sep=comma] {assets/results/f1_scores_adversarial.csv};
\addlegendentry{MIL Attention Zero Padding};

\addplot[color=red, mark=square*] table [x=Epoch, y=MIL Attention Noise Padding, col sep=comma] {assets/results/f1_scores_adversarial.csv};
\addlegendentry{MIL Attention Noise Padding};

\addplot[color=orange, mark=*] table [x=Epoch, y=Base CNN Zero Padding, col sep=comma] {assets/results/f1_scores_adversarial.csv};
\addlegendentry{Base CNN Zero Padding};

\addplot[color=orange, mark=square*] table [x=Epoch, y=Base CNN Noise Padding, col sep=comma] {assets/results/f1_scores_adversarial.csv};
\addlegendentry{Base CNN Noise Padding};

\end{axis}
\end{tikzpicture}

%% file: sections/5_limitations.tex
\subsection{Robustness}
The proposed solution still struggles with noise-enlarged samples, and would therefore likely also struggle with other forms of adversarial enlargement, such as appending malware samples from other families or benign samples to the end or throughout the binary for targeted misclassification attacks. Additionally, although the MIL model performs better than the baseline models in the adversarial setting, an F1 score of 0.705 is quite low. However, further optimisation of the model architecture and hyperparameters could help alleviate these issues, and should be explored in future work.

\subsection{Cost of Full-sized Byteplots}

The results show two main drawbacks of using full-sizes byteplots; longer training times and the addition of noise to the input. Therefore, although the use of full-sized byteplots reduces the effectiveness of enlargement attacks, it comes at the expense of increased costs and model complexity requirements. In the case of malware detection, these trade-offs might be acceptable.

\subsection{Implementation of Models}
The implementation of the models was not optimal, and common techniques such as batch normalisation or normalising image inputs during pre-processing could have been added to improve performance and efficiency. Batch normalisation can improve training performance by addressing internal covariate shift through normalising feature maps over each batch \cite{Ioffe2015BatchShift}, while image normalisation can lead to faster convergence and more stable gradients. 

Furthermore, it could be argued that there is little point in reshaping a one-dimensional input into a two-dimensional matrix and then using a 2D convolution, if instead a 1D convolution could be used directly with the original input. Raff et al. \cite{Raff2017MalwareEXE} mentioned this while discussing alternative architectures for their popular MalConv model (a 1D CNN model). They postulate that a 2D convolution of a byteplot is equivalent to a dilated 1D convolution, and it would perform less well as vertically adjacent pixels in byteplots are not as closely related as they are in natural images. However, their model was developed for binary malware/benign classification, not malware family classification. Malware family classification is more about the structural differences and similarities between samples rather than small byte patterns, and therefore I would argue that 2D convolution is still a useful approach and works well in practice. Although binary executables are naturally sequences, there are clearly both one-dimensional patterns and two-dimensional patterns. Future work could compare 1D convolution versus 2D convolution with MIL for malware family classification. 

\subsection{Imbalanced Learning}
The experiments in this project did not address the data imbalance present in the Microsoft Malware Classification dataset. The dataset only includes 43 samples from the Simda family, while there are 2942 samples from the Kelihos\_ver3 family. As supervised machine learning models are dependent on the quality of their training data \cite{Cortes1995LimitsQuality}, datasets should be representative and balanced to avoid the model becoming biased towards predicting the classes with more samples. Methods such as class weights, oversampling minority classes, and undersampling majority classes to improve performance could have been investigated.

%% file: sections/6_conclusion.tex
This paper investigated adversarial enlargement attacks against visualisation-based malware classification models and proposed a method to overcome these attacks using multiple instance learning. The resizing operations that are commonly used in practice and the literature were proved to be vulnerable, and the proposed solution was shown to achieve a much higher accuracy at classifying adversarial malware samples. 

The results indicate that resizing-based models are able to accurately classify malware families, but when tested with adversarially enlarged malware, they display up to a drastic 97.8\% decrease in performance (F1 score). On the other hand, multiple instance learning models with attention-based instance aggregation show similar performance to the resizing-models for classifying standard malware samples, while only yielding between a 7.7\% and 24.4\% decrease in performance with adversarial examples. The MIL models are able to somewhat overcome the adversarial enlargement attacks, while the baseline models are vulnerable. However, this comes at the cost of an increase in training time and inference time. Although the results of applying the MIL paradigm to this kind of malware classification are promising, there is still much room for improvement.

One of the main benefits of image-based malware classification is the speed, which allows for efficient large-scale classification. Although the proposed MIL method is slower than the resizing-based methods, it is still faster than most static and dynamic methods, taking only 4.67 milliseconds to classify a standard malware sample. Additionally, the other benefits of image-based classification persist, such as traditional obfuscation resistance and platform independence. While static signatures have an inability to generalise \cite{Fleshman2018StaticAnti-Virus} and dynamic analysis can suffer from discrepancies between training data and real-life environments \cite{Rossow2012PrudentOutlook}, classifying byteplots is an alternative method that does not face the same issues.

We conclude that image-based malware classification is most useful for quick and efficient triaging of large quantities of malware samples, but, as with every machine learning solution, has its own weaknesses and should not be used on its own. Complementing image-based techniques with other static and dynamic techniques that have more technical specificity and detail would result in a more complete system. 

The proposed MIL architecture shows promising performance at overcoming adversarial enlargement attacks, and can also be used for other domains to deal with large images of varying size. However, adversarial enlargement attacks are quite rare, and therefore the costs of using the MIL model may outweigh its benefits. In a situation where data loss from resizing would be too extreme, such as in the work from Microsoft and Intel \cite{Chen2020STAMINA:Classification}, the benefits of the proposed method would be more significant. 

\subsection{Future Work}

For future work, it would be worthwhile to investigate the effectiveness of the proposed MIL architecture against other adversarial techniques (targeted misclassification \cite{Khormali2019COPYCAT:Detection}, intra-sample padding enlargement \cite{Benkraouda2021AttacksExecutability}), threat model scenarios (white-box, black-box), and highly obfuscated malware. Additionally, investigating the effectiveness of transfer learning with attention MIL aggregation for classifying adversarially enlarged malware could be interesting.